\documentclass[aps,prl,reprint,superscriptaddress,amsmath,amssymb,pra,showkeys]{revtex4-2}

\usepackage{graphicx}
\usepackage{dcolumn}
\usepackage{bm}
\usepackage{hyperref}
\usepackage{braket}
\usepackage{appendix}
\usepackage{float}
\usepackage{booktabs}
\usepackage{adjustbox}
\usepackage{multirow}
\usepackage{array}
\usepackage{siunitx}
\usepackage{xr}
\usepackage[sort&compress]{cleveref}
\newcolumntype{L}{>{$}l<{$}}
\newcolumntype{C}{>{$}c<{$}}
\newcolumntype{R}{>{$}r<{$}}

\usepackage[hang,flushmargin,symbol*]{footmisc}
\DefineFNsymbolsTM{otherfnsymbols}{
	\textdagger    \dagger
}

\begin{document}
	
	\title{Bichromatic Rabi Control of Semiconductor Qubits}

	\author{Valentin John}
	\thanks{These authors contributed equally to this work.}
	\author{Francesco Borsoi}
	\thanks{These authors contributed equally to this work.}
	\affiliation{QuTech and Kavli Institute of Nanoscience, Delft University of Technology, P.O. Box 5046, 2600 GA Delft, The Netherlands}
	\author{Zoltán György}
	\thanks{These authors contributed equally to this work.}
	\affiliation{Institute of Physics, Eötvös University, H-1117 Budapest, Hungary}
	
	\author{Chien-An Wang}
	\affiliation{QuTech and Kavli Institute of Nanoscience, Delft University of Technology, P.O. Box 5046, 2600 GA Delft, The Netherlands}
	\author{Gábor Széchenyi}
	\affiliation{Institute of Physics, Eötvös University, H-1117 Budapest, Hungary}
	\author{Floor van Riggelen}
	\author{William I. L. Lawrie}
	\author{Nico W. Hendrickx}
	\affiliation{QuTech and Kavli Institute of Nanoscience, Delft University of Technology, P.O. Box 5046, 2600 GA Delft, The Netherlands}
	\author{Amir Sammak}
	\affiliation{QuTech and Netherlands Organisation for Applied Scientific Research (TNO), Stieltjesweg 1, 2628 CK Delft}
	\author{Giordano Scappucci}
	\affiliation{QuTech and Kavli Institute of Nanoscience, Delft University of Technology, P.O. Box 5046, 2600 GA Delft, The Netherlands}
	
	\author{András Pályi}
	\thanks{These authors jointly supervised this work.}
	\affiliation{Department of Theoretical Physics, Institute of Physics, Budapest University of Technology and Economics, Műegyetem rakpart 3, H-1111 Budapest, Hungary}
	\affiliation{MTA-BME Quantum Dynamics and Correlations Research Group, Budapest University of Technology and Economics, Műegyetem rakpart 3, H-1111 Budapest, Hungary}
	
	\author{Menno Veldhorst}
	\thanks{These authors jointly supervised this work.}
	\affiliation{QuTech and Kavli Institute of Nanoscience, Delft University of Technology, P.O. Box 5046, 2600 GA Delft, The Netherlands}
	\email{M.Veldhorst@tudelft.nl}
	\date{\today}            
	
	\begin{abstract}
		Electrically driven spin resonance is a powerful technique for controlling semiconductor spin qubits. 
		However, it faces challenges  in qubit addressability and off-resonance driving in larger systems. 
		We demonstrate coherent bichromatic Rabi control of quantum dot hole spin qubits, offering a spatially selective approach for large qubit arrays. 
		By applying simultaneous microwave bursts to different gate electrodes, we observe multichromatic resonance lines and resonance anticrossings that are caused by the ac Stark shift. 
		Our theoretical framework aligns with experimental data, highlighting interdot motion as the dominant mechanism for bichromatic driving.
		
	\end{abstract}
	
	\maketitle
	
	\section{\label{sec:intro}Introduction}
	Spin qubits based on semiconductor quantum dots represent a promising platform for quantum computing owing to their small qubit footprint, long coherence times, and compatibility with semiconductor manufacturing techniques~\cite{Chatterjee2021SemiconductorPractice, Burkard2023}. However, the present control of spin qubit devices relies on a brute force approach, where each qubit is individually connected to at least one control line. 
	This approach poses a significant challenge for scaling to larger systems and is affecting current progress~\cite{Vandersypen2017InterfacingCoherent, Franke2019RentsComputing}. 
	Multiplexing strategies will most likely become essential and this has been the motivation for various proposals, such as crossbar arrays with shared control~\cite{Li2018, Borsoi2023}.
	Executing quantum algorithms requires selective quantum control, but its implementation in large qubit arrays poses significant challenges.
	Recently, the concept of bichromatic spin resonance has been proposed to offer addressable microwave control in qubit crossbar architectures~\cite{Gyorgy2022}. 
	In this approach, two microwave tones with frequencies $f_\mathrm{w}$ and $f_\mathrm{b}$ are utilized. 
	These tones are applied to the word and the bit line of the crossbar array, respectively. By 	exploiting the nonlinearity of electric dipole spin resonance (EDSR)~\cite{Coish2007, Laird2009, Rashba2011, Scarlino2015,  Romhanyi2015, Takeda2018, Undseth2022, Gilbert2023, Bosco2023}, rotations of a qubit with Larmor frequency of $|f_\mathrm{w} \pm f_\mathrm{b}|$ at the intersection of the two lines [Fig. \ref{fig:device}(a)] can be targeted.
	This operation scheme presents new opportunities for both spatially selective qubit addressing and gate parallelization~\cite{Gyorgy2022}.
	Analogous two-photon processes have been utilized in Rydberg-atom processors~\cite{Levine2018, Ebadi2021} and superconducting qubits~\cite{Valery2022} to optimize qubit performance.	
	
	Here, we investigate experimentally and theoretically the bichromatic driving of semiconductor spin qubits in a 2-qubit system defined in a strained germanium quantum well. 
	We find that both qubits can be coherently driven by mixed frequency signals, including the sum and difference of the corresponding frequencies. 
	We investigate the occurrence of resonance anticrossings in EDSR spectroscopy maps, which originate from the Autler-Townes (also known as ac Stark) shift of a photon-dressed spin transition.
	Additionally, we introduce a model that reveals the importance of spin-conserving and spin-flip tunneling terms in bi- and monochromatic EDSR.

	\section{\label{sec:results}Results}
	\begin{figure}[htp!]
		\centering
		\includegraphics{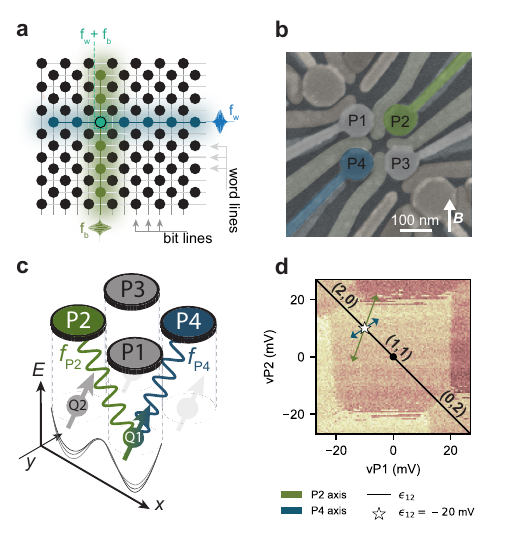}
		\caption{Bichromatic control of a spin qubit. 
			(a) Bichromatic driving in a crossbar architecture.
			(b) False-color scanning electron microscopy of a $2\times2$ germanium quantum dot device, nominally identical to the one used here.
			(c) Illustration of the 4-qubit processor.
			We operate $Q1$ and $Q2$ with microwave bursts applied to $P2$ and $P4$.
			We model qubit rotations via ac detuning modulation (sketched potential).
			(d) Charge stability diagram of the double quantum dot illustrating the $(1, 1)$ charge sector and the detuning $\epsilon_{12}$ axis (black line).			
			The white star indicates the gate voltages used for the qubit manipulation stage. 
			The green and blue arrows indicate the displacement within the $vP1$, $vP2$ framework, when applying a microwave burst on $P2$ and $P4$, showcasing the different orientation of the driving fields. 
			The displayed length of the arrows is proportional to the amplitude of the signal at the device, amplified by a factor of 5 for visibility.		
   		}
		\label{fig:device}
	\end{figure}
	We investigate bichromatic driving of spin qubits in a 2-qubit system within a 4-qubit germanium quantum processor [Figs. \ref{fig:device}(b) and \ref{fig:device}(c)]~\cite{Hendrickx2021AProcessor, Lodari2021}.
	By tuning the electrostatic potential using plunger and barrier gates, we confine a single-hole quantum dot underneath each of the four plungers $P1$-$P4$, and define virtual gate voltages $vP1-vP4$ based on $P1-P4$ to achieve independent control (Supplemental Material Note 1\cite{Supplementary}). 
	We focus on the spin qubits $Q1$ and $Q2$, while $Q3$ and $Q4$ remain in their ground state. 
	We furthermore define the detuning voltage $\epsilon_{12} = vP1 - vP2$~\cite{Hensgens2017}. 
	
\begin{figure*}[htp!]
		\centering
		\includegraphics{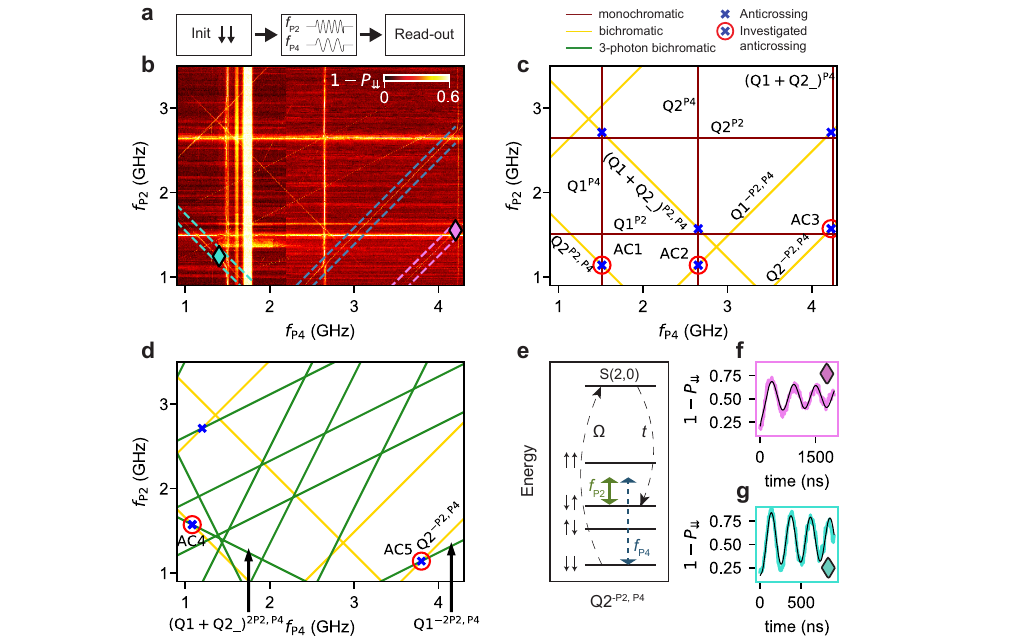}
		\caption{\textbf{Bichromatic EDSR spectroscopy.} 
			\textbf{a,}~Bichromatic control sequence.
			\textbf{b,}~Single-shot probability versus $f_{\mathrm{P4}}$ and $f_{\mathrm{P2}}$, at $\epsilon_{12}= -20$ mV.			  
			We include three turquoise, blue and purple dotted lines to enclose the bichromatic resonances of $\mathrm{Q2^{P2,P4}}$, $\mathrm{Q1^{-P2,P4}}$ and $\mathrm{Q2^{-P2,P4}}$ respectively.
			The broad vertical excitation at $f_{\mathrm{P4}} \sim 1.8$ GHz is associated to a transmission resonance in the lines, and not to a spin transition.		
			\textbf{c, d}~Monochromatic, bichromatic and three-photon bichromatic excitations in the 2D frequency plane, as predicted by theory. 
			\textbf{e,}~Energy diagram of a two-spin system with finite exchange and finite magnetic field. 
			The green and blue arrows represent the applied microwave frequencies $f_{\mathrm{P2}}$ and $f_{\mathrm{P4}}$, when driving the  $\mathrm{Q2^{-P2,P4}}$ transition. 
			Driven spin-flipping processes originate from higher order processes via the S(2,0) state involving the spin-conserving tunneling term $t$ and spin-flip tunneling term $\Omega$.
   			\textbf{f, g} Coherent Rabi oscillations of the $\mathrm{Q2^{-P2,P4}}$ and $\mathrm{Q2^{P2,P4}}$ bichromatic transition. The corresponding $f_{\mathrm{P2}}$ and $f_{\mathrm{P4}}$ frequencies are indicated with the purple and turquoise diamonds in \textbf{b}.
   }
		\label{fig:spectroscopy}
	\end{figure*}

	\autoref{fig:device}(d) displays the charge stability diagram of the double quantum dot system, obtained through rf-reflectometry charge sensing~\cite{Lawrie2020Benchmarks}. 
	The device is operated in an in-plane magnetic field of 0.675 T, resulting in qubit frequencies of $f_{\mathrm{Q1}} = 1.514 $ and $f_{\mathrm{Q2}} = 2.649$ GHz.
	To investigate the bichromatic driving approach, we follow the pulse protocol outlined in Fig. \ref{fig:spectroscopy}(a).
	We initialize the $Q1$, $Q2$ qubits in the $ \ket{\downarrow \downarrow }$ state by adiabatically pulsing $\epsilon_{12}$ from the $(0, 2)$ to the $(1, 1)$ charge state via the spin-orbit induced anticrossing. 
	Next, we manipulate the spins by two simultaneous microwave pulses on plunger gates $P2$ and $P4$, with a duration $t_\mathrm{p}$ and microwave frequencies $f_{P2}$ and $f_{P4}$.
	We perform such two-tone qubit manipulation at the voltage point indicated in Fig. \ref{fig:device}(c) corresponding to $\epsilon_{12}= -20$ mV. 
	Finally, we return to the $(0, 2)$ charge sector and perform readout using latched Pauli spin blockade~\cite{Hendrickx2021AProcessor}.
	
	The 2D EDSR spectroscopy in Fig. \ref{fig:spectroscopy}(b) reveals resonance lines from mono- and bichromatic spin excitations. 
	Monochromatic qubit transitions labeled as $Q1^{P2}$, $Q1^{P4}$, $Q2^{P2}$, $Q2^{P4}$ (with the superscript defining the driving plunger gate) are observed as vertical and horizontal lines at the corresponding Larmor frequencies. 
	Bichromatic excitations appear as tilted resonance lines, with negative (positive) slopes indicating the frequency sum (difference) matching the qubit Larmor frequency.
    Three-photon bichromatic excitations can also be observed when a combination of two photons with the same frequency, and a third one with a different frequency match the qubit Larmor frequency.
	
	Figures \ref{fig:spectroscopy}(c) and \ref{fig:spectroscopy}(d) depict the expected resonance lines considering the individual resonance frequencies of the two qubits. 
	The qubits exchange interaction resulting from interdot tunneling (55 MHz at $\epsilon_{12} = -20$ mV, see Supplemental Material Note 4) is taken into account.
	To label the Larmor frequency of qubit $i$ when qubit $j$ is in the excited state, we use the notation $Qi\_$  (with $i, j \in \{1, 2\}$ and $i \neq j$). 
	The monochromatic transition from $\ket{\downarrow\downarrow}$ to $\ket{\uparrow\uparrow}$ driven by $P4$ is then denoted as $(Q1 + Q2\_ )^{P4}$.
	A bichromatic transition can be visualized as a two-step process via a virtual state, as illustrated in Fig. \ref{fig:spectroscopy}(e). 
	Following perturbation theory, bichromatic spin transitions are activated thanks to spin-conserving ($t$) and spin-flipping ($\Omega$) tunneling terms, which hybridize the four possible spin states with the $S(2,0)$ state, as discussed below and in Supplemental Material Note 8. 
	
	We analyze three resonance lines [dashed lines in Fig. \ref{fig:spectroscopy}(b)] resulting from bichromatic rotation of $Q1$ and $Q2$. 
	The bichromatic $Q1$ spin resonance ($Q1^{-P2,P4}$) occurs when the frequency difference matches the $Q1$ Larmor frequency. 
	Similarly, $Q2$ exhibits bichromatic resonance lines from both frequency difference ($Q2^{-P2,P4}$) and frequency sum ($Q2^{P2,P4}$). 
	The bichromatic spin resonance $Q1^{P2, P4}$ is not investigated due to the presence of a high-pass filter (Supplemental Material Note 6). 
	The conditions for the three studied bichromatic qubit rotations are:  $Q1^{-P2, P4}:  f_{P4}-f_{P2}=f_{Q1}$, $Q2^{-P2, P4}:	f_{P4}-f_{P2}=f_{Q2}$ and $Q2^{P2, P4}: f_{P4}+f_{P2}=f_{Q2}$. 
	At these frequency combinations, we also achieve coherent bichromatic qubit rotations with a Rabi frequency exceeding 1 MHz, as we demonstrate in Figs.~\ref{fig:spectroscopy}(f), \ref{fig:spectroscopy}(g) and Supplemental Material Fig. 4.

	At the intersection of specific resonance lines [see Fig.~\ref{fig:spectroscopy}(b)], we also observe anticrossings labeled as AC$n$ with $n \in \{1,\dots,5\}$ in Figs.~\ref{fig:spectroscopy}(c) and (d).
	\begin{figure*}[htp!]
		\centering
		\includegraphics{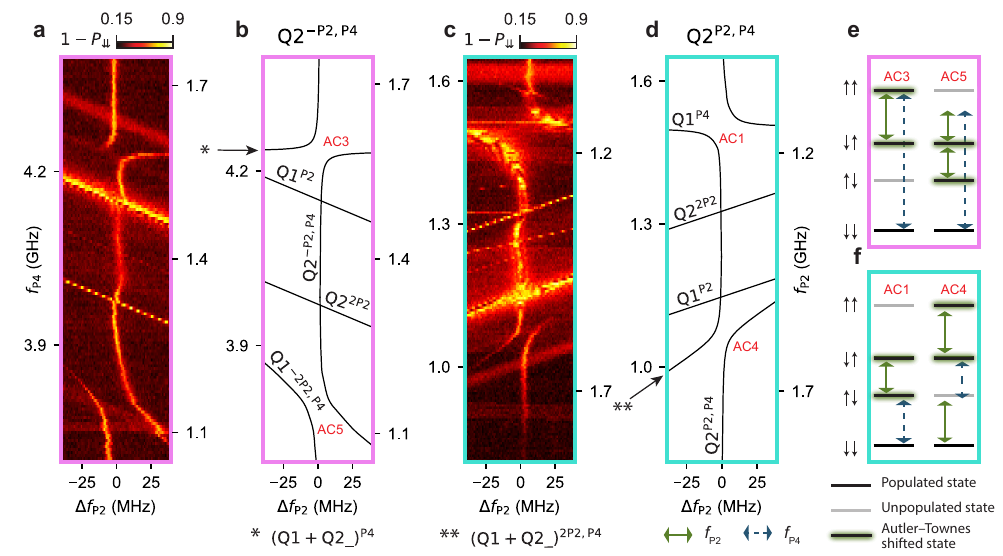}
		\caption{\textbf{Modelling the frequency anticrossings due to the Autler-Townes effect.}
			\textbf{a, c}~Single-shot probabilities ($1 - P_{\downarrow \downarrow}$) in a frequency range around the bichromatic $\mathrm{Q2^{-P2, P4}}$ and $\mathrm{Q2^{P2, P4}}$ resonance conditions, respectively. 
			These scans are higher-resolution measurements along the color-coded diagonals enclosed by two dashed lines in~\autoref{fig:spectroscopy}b. 
			Vertical lines of \autoref{fig:spectroscopy}b appear horizontal, and horizontal lines appear slightly tilted (as can be seen with  $\mathrm{Q1^{P2}}$ and $\mathrm{Q1^{P4}}$ in ~\textbf{d}).     
			The values on the $f_{\mathrm{P2}}$ axes are valid at $\Delta f_{\mathrm{P2}}=0$.
			\textbf{b, d}~Calculated transitions nearby the $\mathrm{Q2^{-P2, P4}}$ and $\mathrm{Q2^{P2, P4}}$ resonances.
			\textbf{e, f}~Illustration of the driven transitions at the four anticrossings. 
			Strong driving via P2 induces a photon-dressed spin transition.
		}
		\label{fig:figure3}
	\end{figure*}
	In Fig.~\ref{fig:figure3}, we analyze the evolution of the two bichromatic spin resonances, $Q2^{-P2, P4}$ and $Q2^{P2, P4}$, in the frequency plane.
	We vary the two microwave frequencies together to follow the two resonance lines, using  $\Delta f_{P2}$ in the range of $[-40, 40]$ MHz centered around the bichromatic resonance. 
	This procedure allows one to monitor in detail the $Q2$ bichromatic spin resonance within the boxed areas indicated in Fig.~\ref{fig:spectroscopy}(b). 
	The bichromatic resonance aligns with the expected value of $\Delta f_{P2} = 0$ for most of the frequency range. 
	However, significant anticrossings occur when the resonance intersects with other qubit transitions.
	Examples of these anticrossings are observed at specific frequencies, and are labeled as AC5, AC3 (for $Q2^{-P2, P4}$), and AC4, AC1 (for $Q2^{P2, P4}$).
	
	The appearance of anticrossings in the frequency plane, such as AC3 in Fig.~\ref{fig:figure3}(a), result from resonant driving of mono- and bichromatic transitions from the $\ket{\downarrow\downarrow}$ state to higher $(1,1)$ states. 
	As shown in Figs.~\ref{fig:figure3}(e) and \ref{fig:figure3}(f), AC3 involves three resonant processes: the bichromatic transition $\ket{\downarrow \downarrow} \leftrightarrow \ket{\downarrow \uparrow}$, the monochromatic $P4$ drive $\ket{\downarrow \downarrow} \leftrightarrow \ket{\uparrow \uparrow}$, and the monochromatic $P2$ drive $\ket{\downarrow \uparrow} \leftrightarrow \ket{\uparrow \uparrow}$. 
	Because of the greater driving efficiency of $P2$ compared to $P4$ [see projected amplitudes in Fig.~\ref{fig:device}(c)], the dominant transition is $\ket{\downarrow \uparrow} \leftrightarrow \ket{\uparrow \uparrow}$ (Supplemental Material Note 8). 
		
	Driving via $P2$ dresses up the spin states $\ket{\downarrow\uparrow}$ and $\ket{\uparrow\uparrow}$ with microwave photons. 
	In the rotating frame where these states are degenerate in the absence of $P2$ driving, the eigenstates become dressed in the form $\frac{\ket{\downarrow\uparrow} \pm  \ket{\uparrow\uparrow}}{\sqrt{2}}$, and the corresponding eigenvalues exhibit a splitting set by the Rabi frequency.
	In this context, dressing refers to the coherent interaction between the electromagnetic field and the spin system, resulting in entangled states of spins and photons becoming the eigenstates of the coupled system. 
	
	This effect, known as the Autler-Townes effect or ac Stark shift, has been observed in quantum optics and in strongly driven superconducting qubits~\cite{Baur2009, Sillanpa2009}. It is at the basis of control strategies for highly coherent solid-state qubits~\cite{Hansen2022}. In particular, the continuous driving can decouple the spin from background magnetic field noise and thus extend their coherence~\cite{Laucht2017, Miao2020}.
	
	Because of the Autler-Townes effect, the resonance frequencies of the two weaker transitions ($\ket{\downarrow \downarrow} \leftrightarrow \ket{\downarrow \uparrow}$ and $\ket{\downarrow \downarrow} \leftrightarrow \ket{\uparrow \uparrow}$) are shifted by the Rabi frequency of the strongly driven $\ket{\downarrow \uparrow} \leftrightarrow \ket{\uparrow \uparrow}$ transition, resulting in the anticrossing between the resonance lines [AC3 in Figs.~\ref{fig:figure3}(a), \ref{fig:figure3}(b)].
	
	We use a two-spin qubit Hamiltonian to model our system and gain a quantitative understanding. 
	The model considers the lowest orbital in each dot, including four states in the $(1,1)$ charge regime, as well as the $(0,2)$ and $(2,0)$ singlet states. 
	Spin-conserving and spin-flip tunneling between the quantum dots are also included, with a coupling strength of $t$ for spin-conserving transitions and $\Omega$ for spin-flip transitions (Supplemental Material Note 8A). 
	Despite neglecting additional electrical $g$-tensor modulations~\cite{Burkard2023,Martinez2022}, this minimal model successfully explains electrically driven spin transitions via ac modulation of the detuning voltage using both mono- and bichromatic resonance techniques. 
	Here, spin dynamics occur through virtual transitions between the $(1,1)$ spin states and the $(0,2)$ and $(2,0)$ singlet states, mediated by the spin-conserving and spin-flipping terms, as shown in Fig.~\ref{fig:spectroscopy}(e).
	
	Our model provides an explanation for the observed resonance crossings and anticrossings in Figs.~\ref{fig:figure3}(a) and \ref{fig:figure3}(c). 	
	Furthermore, by analyzing the five anticrossings AC1$-$AC5 using Floquet theory, as discussed in Supplemental Material Note 8C, we estimate the spin-conserving and spin-flip tunneling energies to be on average $t = (18.1 \pm 1.9)$ and $\Omega = (14.3 \pm 2.4)$ $\SI{}{\micro\electronvolt}$ (Supplemental Material Note 8F).
	
	\begin{figure}[htp!]
		\centering	\includegraphics{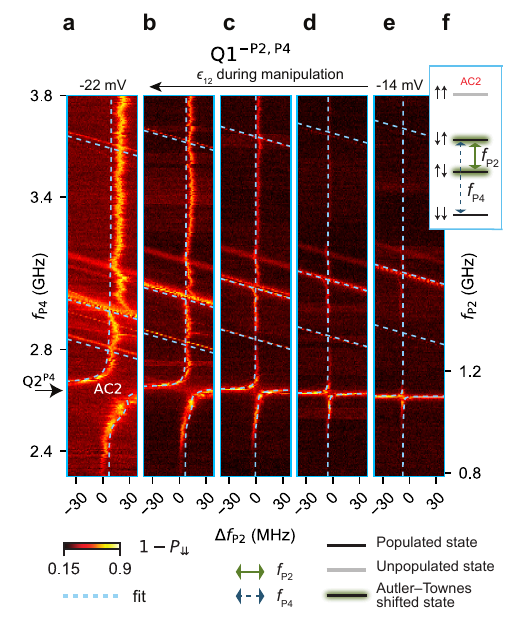}
		\caption{\textbf{Detuning dependence of the frequency anticrossings.} 
			\textbf{a-e,}~Bichromatic spectroscopy around the $f_{\mathrm{Q1}} = f_{\mathrm{P4}} - f_{\mathrm{P2}}$ resonance versus detuning voltage.
			The anticrossing AC2 originates from strong driving of the $\ket{\uparrow\downarrow} \rightarrow \ket{\downarrow\uparrow}$ transition with P2 via the Autler-Townes shift.
			The AC2 frequency gap narrows down as a function of detuning voltage due to suppressed virtual transition from the (1, 1) to the (2, 0) charge state. 
			We overlay the transition lines expected from theory.
			\textbf{f,} Driven transitions at AC2, displaying the four lowest states from \autoref{fig:spectroscopy}c. 
		}
		\label{fig:Q1_difference_detuning}
	\end{figure}
	To verify our theoretical description, we investigate the dependence of the $Q1^{-P2,P4}$ resonance anticrossing on the detuning voltage. 
	Experimental data and the expected detuning dependence from the model are shown in Fig.~\ref{fig:Q1_difference_detuning}. 
	In the model, we use the average tunneling amplitudes and vary the detuning voltage. 
	Moreover, we utilize an estimated detuning lever arm of $\alpha=0.0917$ eV/V and quantum dot charging energy of $U=2.56$ meV (Supplemental Material Note 8B). 
	Our theoretical model accurately captures the diminishing size of the anticrossing as the detuning approaches $\epsilon_{12} \sim 0$. 
	Both the bi- and monochromatic resonance lines fade, indicating a reduced efficiency as detuning approaches zero. This is consistent with our model, since the transitions take place via the $S(0,2)$ and $S(2,0)$ states and in the high detuning limit the transition through $S(0,2)$ dominates the driving. At zero detuning, the two contributions become equal, while the Rabi frequency has a minimum.
 
	The diminished efficiency of bichromatic operations near the charge-symmetry point supports the fundamental role of virtual interdot transitions as the underlying driving mechanism (Supplemental Material Note 3).
	In Supplemental Material Note 8D, we discuss the limitations of our model and suggest that additional mechanisms, such as EDSR induced by $g$-tensor modulation, may be necessary to fully interpret all experimental observations~\cite{Russ2018,Hendrickx2023, Sarkar2023}.
	
	\section{Conclusions}
    Electric dipole spin resonance has enabled high-fidelity quantum gates on individual qubits, but a key challenge is the development of advanced operations for scalable control. Here, we have established bichromatic control of spin qubits, and turned challenges in EDSR~\cite{Undseth2022} into an opportunity for addressable qubit control in larger arrays. 
	Moreover, we showed the relevance of interdot motion in obtaining bi- and monochromatic driving. 
	Furthermore, as the positions of the observed resonance anticrossings are predictable from the qubit parameters, we envision that, while on the one hand these can be exploited for the operation of dressed semiconductor qubits, on the other hand, these frequencies should be avoided when implementing bichromatic EDSR.
	Future experiments may focus on the optimization of bichromatic driving, for example by tuning parameters such as the interdot coupling, aiming to achieve high-fidelity control of large qubit arrays. \\

 All data and analysis underlying this study are available on a 4TU.ResearchData repository~\cite{data_repo}\\
		
	\acknowledgments
	We are grateful to Maximilian Rimbach-Russ, Brennan Undseth and all the members of the Veldhorst laboratory for fruitful discussions.
	We acknowledge support by the Dutch Research Council through an NWO ENW grant and the European Union for an ERC Starting Grant. 
	F.B. acknowledges support from the Dutch Research Council (NWO) via the National Growth Fund program Quantum Delta NL (Grant No. NGF.1582.22.001).
	This research was supported by the Ministry of Culture and Innovation and the National Research, Development and Innovation Office (NKFIH) within the Quantum Information National Laboratory of Hungary (Grant No. 2022-2.1.1-NL-2022-00004), by the ÚNKP-22-1 New National Excellence Program of the Ministry for Culture and Innovation from the source of the National Research Development and Innovation Fund, by NKFIH through the Hungarian Scientific Research Fund (OTKA) Grant No. FK 132146, by the J\'anos Bolyai Research Scholarship of the Hungarian Academy of Sciences, and by the European Union through the Horizon Europe projects Integrated Germanium Quantum Technology (IGNITE) and Quantum Large Scale Integration in Silicon (QLSI).

\bibliography{bibliography} 

\end{document}


\widetext
\begin{center}
	\large \textbf{Supplementary information: \linebreak Bichromatic Rabi control of semiconductor qubits}\\
	
	\normalsize{~\\ Valentin John,$^{1, *}$ Francesco Borsoi,$^{1, *}$ Zoltán György,$^{2, *}$  \\
		Chien-An Wang,$^{1}$ Gábor Széchenyi,$^{2}$ Floor van Riggelen,$^{1}$ \\
		William I. L. Lawrie,$^{1}$ Nico W. Hendrickx,$^{1}$ Amir Sammak,$^{3}$  \\
		Giordano Scappucci,$^{1}$  András Pályi,$^{4, 5, \dagger}$ and Menno Veldhorst $^{1, \dagger}$ }	
	\small{~\\
		$^1$\textit{QuTech and Kavli Institute of Nanoscience, Delft University of Technology, P.O. Box 5046, 2600 GA Delft, Netherlands}\\
		$^2$\textit{ELTE Eötvös Loránd University, Institute of Physics, H-1117 Budapest, Hungary}\\
		$^3$\textit{QuTech and Netherlands Organisation for Applied Scientific Research (TNO), 10 Stieltjesweg 1, 2628 CK Delft, Netherlands}\\
		$^4$\textit{Department of Theoretical Physics, Institute of Physics, Budapest University of Technology and Economics, Műegyetem rakpart 3, H-1111 Budapest, Hungary}\\
		$^5$\textit{MTA-BME Quantum Dynamics and Correlations Research Group, Budapest University of Technology and Economics, Műegyetem rakpart 3., H-1111 Budapest, Hungary}\\			
		$^{*}$ These authors contributed equally \\
		$^{\dagger}$ These authors jointly supervised this work}	
\end{center}

\renewcommand{\thepage}{S\arabic{page}} 
\renewcommand{\thesection}{Supplementary Note \arabic{section}}  

\setcounter{figure}{0}
\setcounter{page}{1}
\setcounter{section}{0}

\renewcommand{\figurename}{\textbf{Supplementary Figure}}
\renewcommand{\thefigure}{\textbf{\arabic{figure}}}
\renewcommand{\tablename}{\textbf{Supplementary Table}}
\renewcommand{\thetable}{\textbf{\arabic{table}}}
\renewcommand{\bibnumfmt}[1]{[S#1]}
\renewcommand{\citenumfont}[1]{S#1}
\renewcommand{\theequation}{\textbf{S\arabic{equation}}}

\section{Virtual gate matrix}
The virtual gate matrix $\boldsymbol{M}$ serves to independently control each quantum dot energy, and has been defined as:
\begin{equation}\label{eq:virtual_plunger}
	\begin{pmatrix}
		\text{vP1} \\ \text{vP2} \\ \text{vP3} \\ \text{vP4}
	\end{pmatrix} 
	=
	\boldsymbol{M}
	\cdot
	\begin{pmatrix}
		\text{P1} \\ \text{P2} \\ \text{P3} \\ \text{P4}
	\end{pmatrix}
	,\qquad
	\boldsymbol{M}
	=
	\begin{pmatrix}
		1.324 & 0.446 & 0.191 & 0.353 \\
		0.483 & 1.231 & 0.396 & 0.234 \\
		0.287 & 0.344 & 1.224 & 0.423 \\
		0.736 & 0.354 & 0.471 & 1.303
	\end{pmatrix} 
\end{equation}
The dc voltages of the two gates at $\epsilon_{12}= 0$ mV (black circle) at the centre of the map shown in Figure 1 are: -1921.3 mV and -1899.0 mV.

\section{Power dependence of anticrossing AC2 ($\mathrm{Q1^{-P2,P4}}$, $\mathrm{Q2^{P4}}$)}

\begin{figure}[H]
	\centering
	\includegraphics{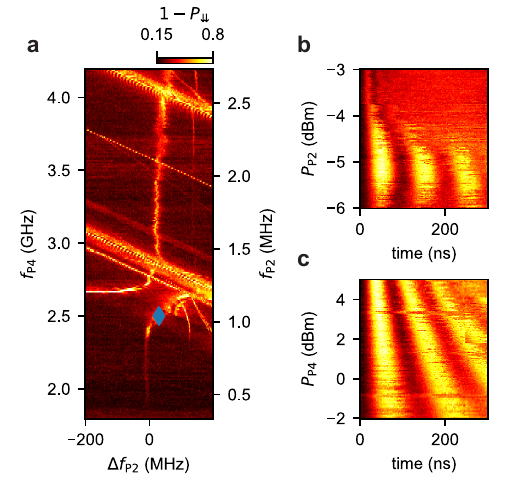}
	\caption{\textbf{Rabi oscillation of $\mathrm{Q1^{P4,-P2}}$ as a function of microwave power applied to the two plunger gates at $\epsilon_{12}= -24$ mV.} 
		\textbf{a,} EDSR spectroscopy map performed at ac driving power of $(P_\mathrm{P2}, P_\mathrm{P4}) = (-5,  3)$ dBm.
		\textbf{b, c} Rabi oscillations as a function of the power on the P2 and P4 driving gates.
  Here, the strong P2 driving field is resonant with the $\uparrow\downarrow \leftrightarrow \downarrow\uparrow$ transition. 
  	This results in a frequency shift of the Lamor frequency of the $\downarrow\downarrow \leftrightarrow \downarrow\uparrow$ transition when sweeping its power. 
  	In contrast, the weak P4 probe field does not alter the Lamor frequency, and sweeping its power leads, in fact, only to an increase in the Rabi frequency. 
  	This behaviour is consistent with the Autler-Townes splitting of the $\uparrow\downarrow$- and $\downarrow\uparrow$-state.
		Both plots are performed with $f_{\text{P4}}=\SI{2.50}{\giga\hertz}$ and $f_{\text{P2}}=\SI{1.07}{\giga\hertz}$.
	}
	\label{fig:power_dep}
\end{figure}

\section{EDSR spectroscopy map at zero detuning}

\begin{figure}[H]
	\centering
	\includegraphics{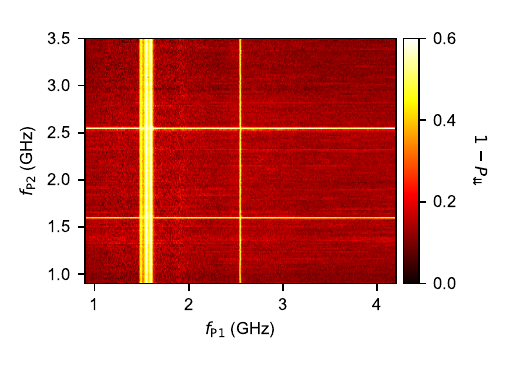}
	\caption{\textbf{Bichromatic EDSR spectroscopy at zero detuning voltage.} 
		We prepare the double quantum dot system in the $\downdownarrows$-state, apply the bichromatic control sequence, and measure the $1-P_\downdownarrows$ probability.
		The major difference to Fig. 2 is the preparation of the system at zero detuning. 
		In this configuration, only two vertical and two horizontal lines are observed, the monochromatic transitions Q1 at $\sim$ 1.60 GHz and Q2 at $\sim$ 2.55 GHz. Note that this plot has been taken with $f_{\text{P1}}$ and $f_{\text{P2}}$, instead of $f_{\text{P2}}$ and $f_{\text{P4}}$. However, we have also observed the absence of the bichromatic signal for $f_{\text{P2}}$ and $f_{\text{P4}}$ when decreasing the detuning.
	} 
	\label{fig:map_zero_det}
\end{figure}

\section{Exchange coupling at operation point}
Here, we determine the exchange coupling at our operation point $\epsilon_{12}= -20$ mV by measuring the resonance frequencies $f_{\mathrm{Q1}}$, $f_{\mathrm{Q1\_}}$, $f_{\mathrm{Q2}}$, and $f_{\mathrm{Q2\_}}$, defined as:
\begin{equation}
	\mathrm{Q1}: \downdownarrows \leftrightarrow \uparrow\downarrow
\end{equation}
\begin{equation}
	\mathrm{Q1\_}: \downarrow\uparrow \leftrightarrow \upuparrows
\end{equation}
\begin{equation}
	\mathrm{Q2}: \downdownarrows \leftrightarrow \downarrow\uparrow
\end{equation}
\begin{equation}
	\mathrm{Q2\_}: \uparrow\downarrow \leftrightarrow \upuparrows
\end{equation}
We compare $f_{\mathrm{Q1}}$ with $f_{\mathrm{Q1\_}}$, and $f_{\mathrm{Q2}}$, $f_{\mathrm{Q2\_}}$, obtaining similar values of 55 MHz for $f_{\mathrm{Q1}}$ and 57 MHz for $f_{\mathrm{Q2}}$.

\begin{figure}[H]
	\centering
	\includegraphics{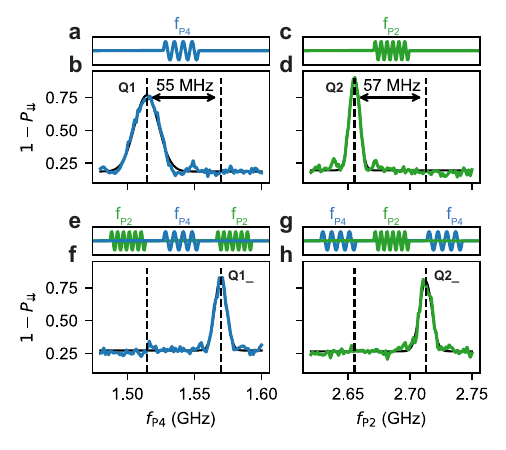}
	\caption{\textbf{Determination of exchange interaction via conditional EDSR spectroscopy. a, c, e, g,} Pulse sequences to prepare the control qubit in the two basis states and probe the target qubit. The blue (green) pulse is applied on P4 (P2) to control Q1 (Q2). \textbf{b, d, f, h} Measurement results from the corresponding pulse sequences to probe Q1, Q1\_, Q2, and Q2\_ respectively. The difference in resonance frequency concludes an exchange interaction of 56 MHz at a voltage of vP1 = -10 mV, and vP2 = 10 mV.
	}
	\label{fig:fres_exchange}
\end{figure}

\section{Resonance line identification and Rabi rotations}
\begin{figure}[H]
	\centering
	\includegraphics{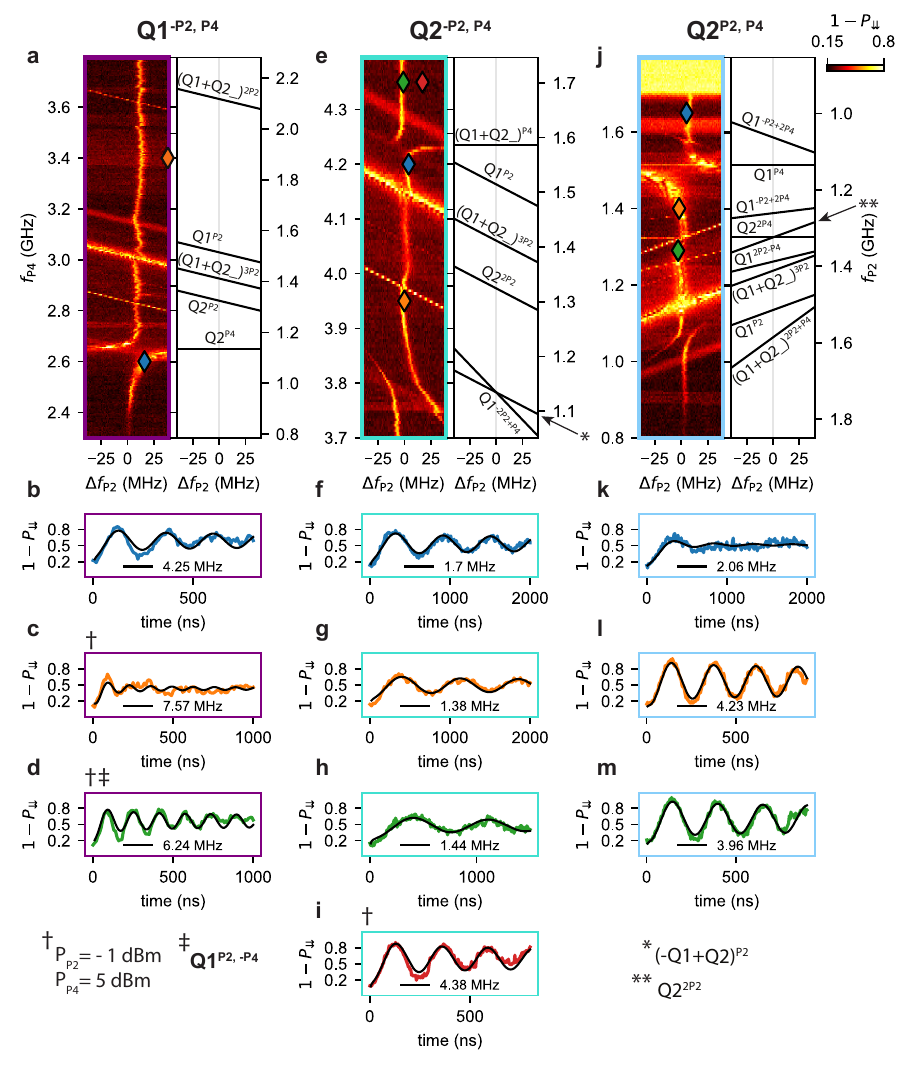}
	\caption{\textbf{Coherent Rabi control by bichromatic driving. a,} EDSR spectroscopy map of $\mathrm{Q1^{-P2,P4}}$ with two indicated positions, where Rabi oscillations have been performed.
		\textbf{b,} Rabi oscillation corresponding to the blue marker in \textbf{a}.
		\textbf{c,} Rabi oscillation corresponding to the orange marker in \textbf{a}.  The marker is shifted 40 MHz away from the bichromatic line because the power on both plungers had to be increased from -5 to -1 dBm for P2 and from 3 to 5 dBm for P4 to see any oscillations. 
		\textbf{d,} Rabi oscillations with similar settings as in \textbf{c}, but the values of $f_\mathrm{{P2}}$ and  $f_\mathrm{{P4}}$
		have been swapped, driving $\mathrm{Q1^{P2,-P4}}$ instead of $\mathrm{Q1^{-P2,P4}}$. 
		Note that the corresponding marker falls outside the depicted regime shown in \textbf{a}.
		\textbf{e,} EDSR spectroscopy map of $\mathrm{Q2^{-P2,P4}}$ with four indicated positions, where Rabi oscillations have been performed.
		\textbf{f, g, h,} Rabi oscillation corresponding to the blue, orange and green marker in \textbf{e}.
		\textbf{i, } same as in \textbf{h}, but with power on both plungers  increased from -5 to -1 dBm for P2 and from 3 to 5 dBm for P4.
		\textbf{j,} EDSR spectroscopy map of $\mathrm{Q2^{P2,P4}}$ with three indicated positions, where Rabi oscillations have been performed.
		\textbf{k, l, m, } Rabi oscillation corresponding to the blue, orange, and green marker in \textbf{j}.}
	\label{fig:anti-crossings}
\end{figure}

To identify all the resonance lines of Figs. 3 and 4, we simulate the position of all the expected lines neglecting the Autler-Townes effect. 
We can write the considered bichromatic transitions as
\begin{equation} \label{Q1_dif_bicon}
	m \cdot f_{\text{P2}}^\textrm{offset} + f_{\text{P4}} = f_{Q}, m = \pm 1,
\end{equation}
where $f_Q$ represents the qubit frequencies $f_{\mathrm{Q1}}$ and $f_{\mathrm{Q2}}$. 
Our expected resonance lines follow
\begin{equation}  \label{Q1_dif_reslines}
	f_\textrm{res} = n_{\text{P2}} \cdot f_{\text{P2}} + n_{\text{P4}} \cdot f_{\text{P4}},
\end{equation}
where $f_{\text{P2}}=f_{\text{P2}}^\textrm{offset} + \Delta f_{\text{P2}}$. 
Here, $f_\textrm{res}$ can be $f_{\mathrm{Q1}}$, $f_{\mathrm{Q2}}$,  or $f_{\mathrm{Q1}}+f_{\mathrm{Q2\_}}$, and $n_{\text{P2}}$ and $n_{\text{P4}}$ are integers referring to the considered harmonic. By plugging in \autoref{Q1_dif_bicon} into \autoref{Q1_dif_reslines}, we obtain
\begin{equation}
	f_{\text{P4}} = \frac{f_\textrm{res}}{m \cdot n_{\text{P2}}+  n_{\text{P4}}} + \frac{n_{\text{P2}}}{m \cdot n_{\text{P2}} + n_{\text{P4}}} \cdot (m \cdot f_{Q1} - \Delta f_{\text{P2}})
\end{equation}
We use these equations to calculate the expected resonance lines as visible in Suppl. Fig.~\ref{fig:anti-crossings}.

\section{Attenuation}

\begin{figure}[hbt]
	\centering
	\includegraphics{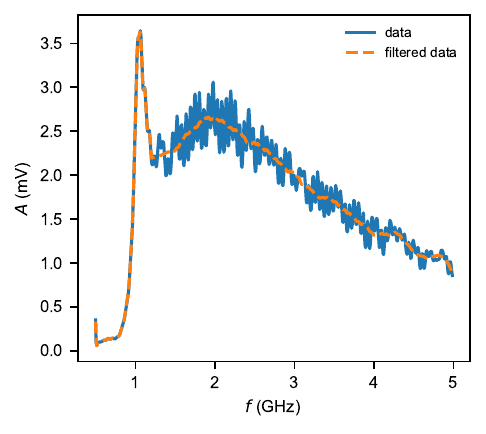}
	\caption{ \label{fig:attenuation} \textbf{Attenuation caused by diplexer and fridge cables.} Amplitude of the signal arriving at the device level considering $P_{\text{P4}} = \SI{2.5}{dBm}$. The signal of P2 with $P_{\text{P2}}= -\SI{6}{dBm}$ is approximately the same since it has 8.5 dB less attenuation on the lines. A Savitzky-Golay filter is applied on the data, the result is shown with orange-dashed line. 
	}
	\label{fig:power_analysis_setup}
\end{figure}

We approximate the amplitude arriving at the device by measuring the attenuation of the fridge lines of a comparable setup at cryogenic temperature, and the frequency response of an equivalent diplexer that is used to combine the microwave signal with the baseband pulses from the AWG.
The measurement data with the Savitzky-Golay filter can be seen in Suppl. Fig.~\ref{fig:attenuation}. Since the measurement has only been performed on a similar setup, this constitutes only an approximation, but the general shape of the attenuation is expected to be the same.

\section{Monochromatic Rabi frequencies}

Suppl. \autoref{tab:power_mono} shows the measured monochromatic Rabi frequencies of the two considered qubits. Notably, $\text{Q1\_}^{\text{P4}}$ differs by a factor of two from  $\text{Q1}^\text{P4}$. $P_{\text{mw}}$ is the microwave power at the output of the signal generator, and corresponds to $P_{\text{P2}}$ for Q1 and $P_{\text{P2}}$ for Q2.

\begin{table}[hbt]
	\renewcommand{\arraystretch}{1.4}
	\caption{}
	\label{tab:power_mono}
	\begin{tabular}{|c|c|c|c|c|c|c|c|}
		\hline
		\textbf{Transition} & 
		\textbf{$f_{\textrm{res}}$ {(}GHz{)}} &
		\textbf{$t_{\textrm{Rabi}}$ {(}ns{)}} &
		\textbf{$f_{\textrm{Rabi}}$ {(}MHz{)}} &
		\textbf{$P_{\textrm{mw}}$ {(}dBm{)}} &
		\textbf{$A$ {(}mV{)}} \\ \hline
		$\text{Q1}^{\text{P4}}$    & 1.51             & 42.5                & 11.76                & 2.5           & 2.3     \\ \hline
		$\text{Q1\_}^{\text{P4}}$  & 1.57             & 88.6                & 5.65                & 2.5           & 2.4     \\ \hline
		$\text{Q2}^{\text{P2}}$    & 2.65             & 24.2                & 20.66                & -6.0          & 2.3      \\ \hline
		$\text{Q2\_}^{\text{P2}}$  & 2.74             & 22.6                & 22.1                & -5.8          & 2.2     \\ \hline
	\end{tabular}
\end{table}

\section{Theoretical analysis}

In this Supplementary Note, we detail the elements of our theoretical analysis. 

\subsection{Model Hamiltonian}
\label{subsec:modelhamiltonian}

Here we describe in details the model Hamiltonian of the two-hole double quantum dot introduced in the main text. The matrix representation of that Hamiltonian reads:
\begin{equation}\label{equ:hamiltonian}
	H=\begin{pmatrix}
		-\hbar\delta\omega_z & 0 & 0 & 0 & i\hbar\Omega_z-t_0 & i\hbar\Omega_z-t_0 \\
		0 & \hbar\delta\omega_z & 0 & 0 & i\hbar\Omega_z+t_0 & i\hbar\Omega_z+t_0 \\
		0 & 0 & \hbar\omega_z & 0 & -i\hbar\Omega_x-\hbar\Omega_y & -i\hbar\Omega_x-\hbar\Omega_y \\
		0 & 0 & 0 & -\hbar\omega_z & i\hbar\Omega_x-\hbar\Omega_y & i\hbar\Omega_x-\hbar\Omega_y \\
		-i\hbar\Omega_z-t_0 & -i\hbar\Omega_z+t_0 & i\hbar\Omega_x-\hbar\Omega_y & -i\hbar\Omega_x-\hbar\Omega_y & U-\epsilon(t) & 0 \\
		-i\hbar\Omega_z-t_0 & -i\hbar\Omega_z+t_0 & i\hbar\Omega_x-\hbar\Omega_y & -i\hbar\Omega_x-\hbar\Omega_y & 0 & U+\epsilon(t) 
	\end{pmatrix}, 
\end{equation}
where the basis states in order are  $\ket{\uparrow, \downarrow}$, $\ket{\downarrow,\uparrow}$, $\ket{\uparrow,\uparrow}$, $\ket{\downarrow,\downarrow}$, $\ket{0,2}$ and $\ket{2,0}$. 
We have defined the symmetric and asymmetric Zeeman splittings, respectively, as 
$\hbar \omega_z = \frac{1}{2} (g_1+g_2) \mu_B B$ and
$\hbar \delta \omega_z = \frac{1}{2} (g_2-g_1) \mu_B B$, 
where $g_1$ and $g_2$ are the g-factors of Q1 and Q2 dot, $\mu_B$ is the Bohr magneton and $B$ is the  magnetic field. The g-factors are assumed to depend linearly on the virtual plunger gate voltages vP1 and vP2: 
\begin{equation}
	g_1=g_{1,0}+A_1\mathrm{vP1}+B_1\mathrm{vP2}, \hspace{5mm}     g_2=g_{2,0}+A_2\mathrm{vP1}+B_2\mathrm{vP2}. 
\end{equation}
The on-site Coulomb repulsion energy is denoted by $U$. The spin-independent interdot tunneling amplitude is described by $t_0$, and spin-dependent tunneling due to spin-orbit interaction is characterised by the vector $\Omega = (\Omega_x,\Omega_y,\Omega_z)$. 

We anticipate that the positions of the monochromatic and bichromatic transitions in the $(f_{\text{P2}}, f_{\text{P4}})$ frequency plane, as well as the resonance anticrossing curves in that plane, can be described in terms of the following two combined tunneling amplitudes: 
\begin{equation}
	t e^{i \Phi} = t_0 + i \hbar \Omega_z, \hspace{5mm} \Omega e^{i \Phi_c} = \hbar (\Omega_y + i \Omega_x),
\end{equation}
where 
$t, \Omega > 0$ and $\Phi, \Phi_c \in [0,2\pi)$. 
We call $t$ and $\Omega$ the spin-probability conserving and spin-probability flipping tunnelings.
The results of our present analysis are insensitive to the phase angles $\Phi$ and $\Phi_c$. 
However, these phase angles are relevant in the sense that they affect the fine structure of the resonance anticrossings, and contribute to interference effects if additional driving mechanisms are taken into account, besides the detuning modulation in $H$.

In Eq.~\eqref{equ:hamiltonian}, the on-site energy difference (\emph{detuning}) is expressed from the virtual plunger gate voltages as: 
\begin{equation}\label{equ:on-site} \epsilon(t)=\epsilon+\delta\epsilon(t)=\alpha(\mathrm{vP1}-\mathrm{vP2}+\delta \mathrm{vP1}(t)-\delta \mathrm{vP2}(t)).
\end{equation}
Here, $\epsilon$ is the static component of the detuning, and $\delta\epsilon(t)$ is the detuning modulation, furthermore, $\alpha=0.0917$ eV/V is the lever arm, and $\delta \mathrm{vP1}(t)$ and $\delta \mathrm{vP2}(t)$ are the ac components of the virtual plunger gate voltages describing the driving signals.
The lever-arm $\alpha$ describes how much the quantum dots' on-site energies are changed by the virtual plunger gate voltages, and the lever arm is the same for plungers vP1 and vP2 with good approximation. 
The modulations of the virtual plunger gate voltages $\delta \mathrm{vP1}(t)$ and $\delta \mathrm{vP2}(t)$ depend on the ac signals on P4 and P2:
\begin{equation}
	\begin{pmatrix}
		\mathrm{vP1}(t) \\ 
		\mathrm{vP2}(t)
	\end{pmatrix}= \begin{pmatrix}
		\boldsymbol{M}_{12} & \boldsymbol{M}_{14} \\ 
		\boldsymbol{M}_{22} & \boldsymbol{M}_{24} 
	\end{pmatrix}
	\begin{pmatrix}
		E_{P2}\cos{\omega_2 t}\\ 
		E_{P4}\cos{\omega_4 t}
	\end{pmatrix},
\end{equation}
where $E_{P2}$ and $E_{P4}$ denote actual plunger gate voltages on P2 and P4. 
Hence, the detuning modulation $\delta\epsilon(t)$ can be written as: 
\begin{equation}\label{eq:delta_epsilon}
	\delta\epsilon(t)=\alpha(\boldsymbol{M}_{12}-\boldsymbol{M}_{22})E_{P2}\cos{\omega_2t}+\alpha(\boldsymbol{M}_{14}-\boldsymbol{M}_{24})E_{P4}\cos{\omega_4 t}=\epsilon_{\text{P2}}\cos{\omega_2t}+\epsilon_{\text{P4}}\cos{\omega_4t},  
\end{equation}
where we introduced $\epsilon_{\text{P2}}=\alpha(\boldsymbol{M}_{12}-\boldsymbol{M}_{22})E_{P2}$ and $\epsilon_{\text{P4}}=\alpha(\boldsymbol{M}_{14}-\boldsymbol{M}_{24})E_{P4}$, while $\boldsymbol{M}_{12}=0.446$, $\boldsymbol{M}_{22}=1.231$, $\boldsymbol{M}_{14}=0.353$ and $\boldsymbol{M}_{24}=0.234$, the corresponding matrix elements in Eq. \ref{eq:virtual_plunger}.  

\subsection{Inference of the parameters of the static Hamiltonian}

\begin{figure}[h]
	\centering
	\includegraphics[width=\columnwidth]{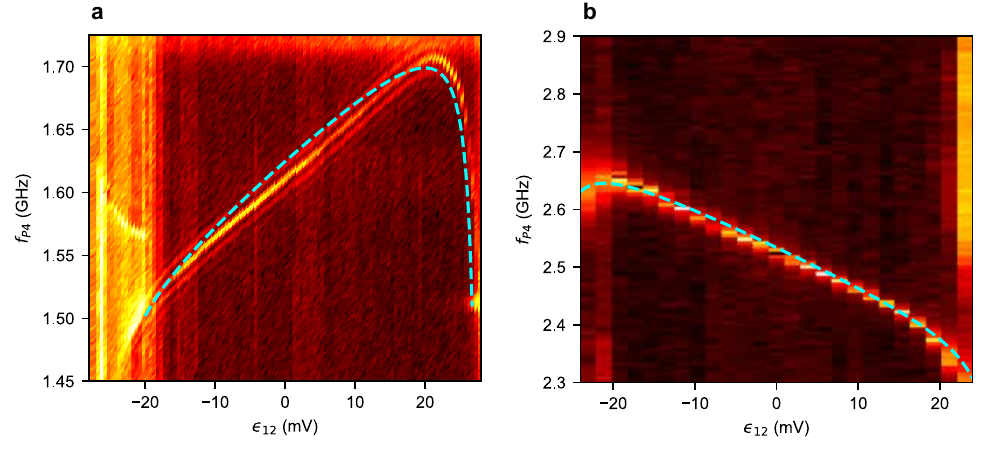}
	\caption {\label{fig:fit_spectrum} \textbf{Inference of the parameters of the static Hamiltonian from EDSR spectroscopy.} 
		\textbf{a.} 
		Experimental EDSR spectroscopy data showing the resonance frequency $f_\mathrm{Q1}$ as function of detuning voltage $\epsilon_{12}$. The pulse sequence consisted of a pulse on P4 with $t_{\text{mw,P4}} = \SI{118}{ns}$ and $P_{P4}=- \SI{1}{dBm}$, where its frequency $f_{\text{P4}}$ has been swept from 1.45 to 1.75 GHz. Before and after this pulse we apply another fixed pulse on P2 with $f_{\text{mw,P2}} = \SI{2.67}{GHz}$, $P_{P2}=- \SI{15}{dBm}$ and $t_{\text{mw,P2}} = \SI{34}{ns}$, which creates a superposition state at $\epsilon_{12} = - \SI{20}{mV}$. 
		Dashed line shows the fit of Eq. \ref{eq:fQ1_fit} on the data. 
		\textbf{b.}
		Experimental EDSR spectroscopy data showing the resonance frequency $f_\mathrm{Q2}$ as function of the detuning voltage $\epsilon_{12}$. The pulse sequence consisted of a pulse on P2 with $t_{\text{mw,P2}} = \SI{118}{ns}$ and $P_{P2}=- \SI{5}{dBm}$, where its frequency $f_{\text{P4}}$ has been swept from 2.3 to 2.9 GHz. No additional pulse has been applied. Dashed line shows the fit of Eq. \ref{eq:fQ2_fit} on the data.}
\end{figure}

Our goal in this subsection is to determine the parameters of the static part of the Hamiltonian $H$ of Eq.~\eqref{equ:hamiltonian}.
To this end, we use experimental EDSR spectroscopy data, taken as the function of the drive frequency $f_{\text{P4}}$ and the detuning along the $\mathrm{vP2} = - \mathrm{vP1}$ line.
The data for the two qubits Q1 and Q2 is shown in Suppl. Fig.~\ref{fig:fit_spectrum}a and b, respectively.
We fit the spectrum of the model Hamiltonian (Eq.~\ref{equ:hamiltonian}) on the resonance frequencies seen in this spectroscopy data.

The fitting procedure is as follows.
We order the spectrum of the static part of $H$ as $E_1 < E_2 < E_3 < E_4 < E_5 < E_6$.
Using second-order Schrieffer-Wolff  transformation
\cite{winkler2003quasi}  on the (1,1) charge subspace, we obtain approximate results for the resonance frequencies $f_\mathrm{Q1}$ and $f_\mathrm{Q2}$:
\begin{equation}\label{eq:fQ1_fit}
	h f_\mathrm{Q1}
	=
	E_2 - E_1
	\approx
	{\mu_B B}\left[g_{1,0}+\dfrac{A_1-B_1}{2}\epsilon_{12}\right]+2\dfrac{\Omega^2-t^2}{ U}\dfrac{1}{1-\frac{\alpha^2}{U^2}\epsilon_{12}^2},
\end{equation}
\begin{equation}\label{eq:fQ2_fit}
	h f_\mathrm{Q2}
	=
	E_3 - E_1
	\approx
	{\mu_B B}\left[g_{2,0}+\dfrac{A_2-B_2}{2}\epsilon_{12}\right]+2\dfrac{\Omega^2-t^2}{ U}\dfrac{1}{1-\frac{\alpha^2}{U^2}\epsilon_{12}^2}.
\end{equation}
Note that both formulas contain four fitting parameters, and two of those fitting parameters are the same. 
We first fit Eq.~\eqref{eq:fQ1_fit} on the resonance curve shown in Suppl. Fig.~\ref{fig:fit_spectrum}a, and hence obtain these values:
\begin{equation}
	g_{1,0}=0.174, \hspace{5mm} A_1-B_1=1.043 \cdot 10^{-3} \hspace{1mm} \frac{1}{\mathrm{mV}}, 
\end{equation}
\begin{equation}\label{eq:fitresult:Ot}
	\dfrac{\Omega^2-t^2}{U}=-0.0474 \hspace{1mm} \mu\mathrm{eV}, \hspace{5mm} \dfrac{\alpha}{U}=0.0358 \hspace{1mm} \frac{1}{\mathrm{mV}}.
\end{equation}
Then, we insert the parameter values in Eq. \eqref{eq:fitresult:Ot} into Eq. \eqref{eq:fQ2_fit}, and infer the remaining two parameters of \eqref{eq:fQ2_fit} by fitting that on the resonance curve of Suppl. Fig.~\ref{fig:fit_spectrum}b:
\begin{equation}
	g_{2,0}=0.271, \hspace{5mm} A_2-B_2=1.426\cdot 10^{-3} \hspace{1mm} 
	\frac{1}{\mathrm{mV}}.
\end{equation}

\subsection{Analytical methods to describe EDSR resonances: Floquet theory and Schrieffer-Wolff transformation}
\label{appsubsec:analytical}

In this section, we introduce and apply Floquet theory and Schrieffer-Wolff perturbation theory as analytical tools to derive formulas for the monochromatic and bichromatic Rabi frequencies, and to describe the shape of the resonance anticrossings shown in Figs. 3 and 4 of  the main text.

The monochromatic and bichromatic Rabi frequencies can be derived via calculating a 2$\times 2$ effective Floquet matrix, while the resonance anticrossings can be described by deriving a $3\times 3$ effective Floquet matrix.
To these ends, we will use the generalization of Floquet theory \cite{shirley1965solution} 
for multiple driving fields, the many-mode Floquet theory \cite{ho1983semiclassical} 
, and will utilize fourth order Schrieffer-Wolff transformation \cite{schrieffer1966relation,winkler2003quasi}.

A key ingredient of Floquet theory is the Floquet matrix, which is derived from the time-dependent Hamiltonian $H(t)$ defined in Eq.~\eqref{equ:hamiltonian}.
For the Floquet analysis, it is useful to separate the static and time-dependent parts of $H$ of Eq.~\eqref{equ:hamiltonian}:
\begin{equation}\label{general_Hamiltonian}
	H(t) =
	H_0+V_2\cos{(\omega_2 t)}+V_4\cos{(\omega_4 t)}, 
\end{equation}
where $H_0$ is the static part, and the operators $V_2$ and $V_4$ of the driving term, corresponding to ac driving via actual plunger gates P2 and P4, can be expressed with the quantities introduced in~\ref{subsec:modelhamiltonian} via Eqs.~\eqref{equ:hamiltonian}, \eqref{equ:on-site}, and \eqref{eq:delta_epsilon}. 

We briefly outline the key elements of Floquet theory as applied for a bichromatically driven system.
Standard Floquet theory provides a solution for the Schr\"odinger equation of a periodically driven quantum system.
In a strict sense, our Hamiltonian $H(t)$ of Eq.~\eqref{general_Hamiltonian} is not periodic in time. 
However, our problem can nevertheless be addressed using standard Floquet theory \cite{ho1983semiclassical}, if we introduce a frequency unit $\omega$ such that
\begin{equation}
	\label{eq:deltaomega}
	\omega_2=N_2\omega, \hspace{3mm} \omega_4= N_4\omega,   
\end{equation}
where $N_2$ and $N_4$ are positive integers.
This frequency unit $\omega$ can be arbitrarily small, this way $\omega_2$ and $\omega_4$ can be written in the form \eqref{eq:deltaomega} with good precision. 
The Hamiltonian is periodic in time if we use this $\omega$ parameter, with time period $T=2\pi/\omega$, which makes standard Floquet theory as described in  \cite{shirley1965solution}.
The problem of the time-dependent Hamiltonian can be transformed into the eigenvalue problem of an infinite-dimensional \emph{Floquet matrix}, which can be obtained by expanding the time-dependent wave function as well as the Hamiltonian in time Fourier series \cite{shirley1965solution}. 

The elements of the Floquet matrix are indexed by the labels $\alpha$ of a basis of the 6-dimensional Hilbert space and the Floquet indices ($n_2,n_4,k_2,k_4 \in \mathbb{Z}$) corresponding to the two drive frequencies.
Below, we use the states of the product basis, as defined below Eq.~\eqref{equ:hamiltonian}; that is, from now on, we have 
$\alpha \in \left\{
\ket{\uparrow, \downarrow}, \ket{\downarrow,\uparrow}, \ket{\uparrow,\uparrow}, \ket{\downarrow,\downarrow},
\ket{0,2}, \ket{2,0}
\right\}$.
Furthermore, Floquet indices are sometimes referred to as photon numbers, due to the strong analogy between Floquet theory and quantum electrondynamics at high photon numbers \cite{shirley1965solution}.
For our problem, the Floquet matrix reads:
\begin{equation}\label{Floquet-matrix}
	\begin{aligned}
		& \mathcal{H}_{F,\alpha n_2 n_4,\beta k_2k_4} \equiv
		\bra{\alpha n_2 n_4}\mathcal{H}_F\ket{\beta k_2 k_4}=\hbar(n_2\omega_2+ n_4\omega_4)\delta_{\alpha\beta}\delta_{n_2k_2}\delta_{ n_4 k_4}+H_{0,\alpha\beta}\delta_{n_2k_2}\delta_{n_4k_4}+\\&+\frac{V_{2,\alpha\beta}}{2}\left(\delta_{n_2-k_2,1}+\delta_{n_2-k_2,-1}\right)\delta_{n_4k_4}+\frac{V_{4,\alpha\beta}}{2}\left(\delta_{n_4-k_4,1}+\delta_{n_4-k_4,-1}\right)\delta_{n_2k_2},
	\end{aligned}
\end{equation}
where 
$H_{0,\alpha\beta}$, $V_{2,\alpha\beta}$ and $V_{4,\alpha\beta}$ denote the matrix elements of $H_0$, $V_2$ and $V_4$. 

In Eq.\eqref{Floquet-matrix}, we have introduced a bra-ket notation for the elements of the Floquet matrix. 
These kets $\ket{\alpha n_2 n_4}$ will be regarded as basis vectors of the Floquet space.
Furthermore, we will refer to these basis vectors $\ket{\alpha n_2 n_4}$ and the corresponding diagonal matrix elements of the Floquet matrix $\braket{\alpha n_2 n_4|\mathcal{H}_F|\alpha n_2 n_4}$ as \emph{Floquet levels}.

The Floquet matrix has a block structure, as seen from  Eq.~\eqref{Floquet-matrix}. 
For each photon number pair $(n_2,n_4)$, there is a $6 \times 6$ block along the diagonal of the Floquet matrix, formed by the static part of the Hamiltonian $H_0$, shifted by $\hbar(n_2\omega_2+n_4\omega_4)$, specified by the first two terms on the right hand side of Eq.~\eqref{Floquet-matrix}. 
These $6 \times 6$ blocks along the diagonal are coupled by the driving term $V_2 \cos{\omega_2 t}+V_4 \cos{\omega_4 t}$, which appear in the Floquet matrix as the last two terms of Eq.~\eqref{Floquet-matrix}.

From the eigenvectors and eigenvalues of the infinite-dimensional Floquet matrix, one can reconstruct the stationary solutions of the time-dependent Schr\"odinger equation. 
In this work, we derive analytical, perturbative approximations for the few relevant eigenvectors and eigenvalues.
We do this using Schrieffer-Wolff perturbation theory, which yields a small effective Floquet matrix of size $2 \times 2$ or $3 \times 3$, for the cases of our interest.
In this perturbative approach, we define the unperturbed Floquet matrix $\mathcal{H}_{F,0}$ as the diagonal part of the Floquet matrix $\mathcal{H}_F$, and the rest (off-diagonal part, including tunneling, spin-orbit interaction and the electric drivings) of the Floquet matrix, $\mathcal{H}_F - \mathcal{H}_{F,0}$, is the perturbation.

\begin{figure}
	\centering
	\includegraphics[width=0.7\columnwidth]{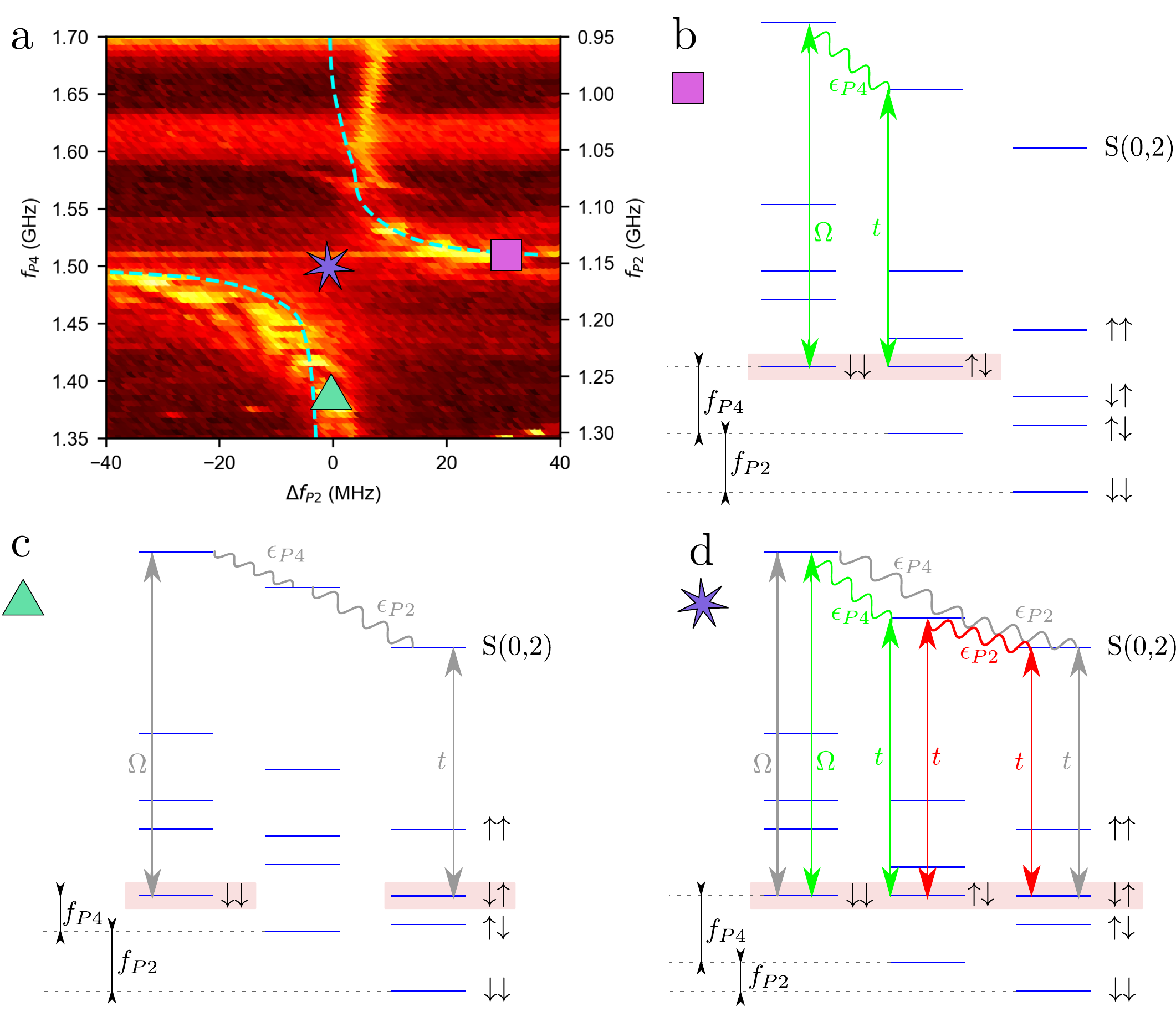}
	
	\caption {\label{fig:floquet-levels} \textbf{Electrically driven monochromatic and bichromatic transitions, and their resonance anticrossing, described using Floquet theory.} 
		\textbf{a.} Measured data of the anticrossing AC1 of the monochromatic single-photon transition $\mathrm{Q1^{P4}}$ and the bichromatic two-photon transition $\mathrm{Q2^{P2,P4}}$. Dashed lines show the result of fitting Eq.~ \eqref{eq:anticrossing_upperQ2sum}. 
		Marked points indicate monochromatic (purple square) and bichromatic (green triangle) transitions, and the centre of the resonance anticrossing (blue star) at the intersection of the monochromatic and bichromatic resonances. 
		\textbf{b.} Floquet level diagram of monochromatic transition $\mathrm{Q1^{P4}}$. The two degenerate Floquet levels are $\ket{\downarrow\downarrow,0,0}$ and $\ket{\uparrow\downarrow,0,-1}$, which are connected by the electric-field matrix element of plunger P4, $\epsilon_{\text{P4}}$, and the two tunnelings $t$ and $\Omega$. 
		\textbf{c.} Floquet level diagram of bichromatic transition $\mathrm{Q2^{P2, P4}}$. The two degenerate Floquet levels are $\ket{\downarrow\downarrow,0,0}$ and $\ket{\downarrow\uparrow,-1,-1}$, which are connected by the two electric-field matrix elements $\epsilon_{\text{P2}}$, $\epsilon_{\text{P4}}$ and the two tunnelings $\Omega$ and $t$. 
		\textbf{d.} Floquet level diagram of the anticrossing of $\mathrm{Q2^{P2, P4}}$ with $\mathrm{Q1^{P4}}$ . The anticrossing is formed when the conditions of both monochromatic and bichromatic transitions are fulfilled, therefore the three Floquet levels mentioned above are degenerate. For simplicity, only half of the pathways through the $S(0,2)$ are shown, but the other half and the ones going through $S(2,0)$ look  similar.
	}
\end{figure}

In the case of resonant monochromatic or resonant bichromatic driving, the resonance condition implies the twofold degeneracy of the relevant Floquet levels. 
These degenerate Floquet levels have different spin configurations, and their photon numbers differ by one or two.  
The Schrieffer-Wolff transformation is done onto this degenerate subspace, to derive an effective $2\times2$ Floquet matrix, in which the absolute value of the off-diagonal matrix element is half of the corresponding Rabi frequency (see below).
Simultaneous monochromatic and bichromatic driving with two simultaneously satisfied resonance conditions can lead to three-fold degenerate Floquet levels; then our Schieffer-Wolff transformation results in a larger, $3\times3$ effective Floquet matrix, whose off-diagonal elements determine if a resonance anticrossing is developed. 

We show examples in Suppl. Fig.~\ref{fig:floquet-levels} for the different driving schemes mentioned in the previous paragraph.
In Suppl. Fig.~\ref{fig:floquet-levels}b, due to an applied resonant monochromatic drive, the Floquet levels $\ket{\downarrow\downarrow,0,0}$, $\ket{\uparrow\downarrow,0,-1}$ are degenerate. 
The Rabi frequency of that transition can be derived using a third-order Schrieffer-Wolff transformation. 
In the case of resonant bichromatic driving, sketched in Suppl. Fig. \ref{fig:floquet-levels}c,  the  Floquet levels  $\ket{\downarrow\downarrow,0,0}$, $\ket{\downarrow\uparrow,-1,-1}$ are degenerate, and the corresponding Rabi frequency can be derived via a fourth-order Schrieffer-Wolff transformation.  

The case with a simultaneous resonance is depicted in Suppl. Fig.~\ref{fig:floquet-levels}d, where three Floquet levels are  degenerate. 
The off-diagonal Floquet matrix elements between these Floquet levels are visualized by different colors. 
Note that although Suppl. Fig.~\ref{fig:floquet-levels}d illustrates each resonant transition as a single process for simplicity, in fact each transition is defined by the interference of multiple processes.

To derive simple formulas for the transition matrix elements, we assume
\begin{equation}
	U\pm\epsilon+n_2\hbar\omega_2+n_4\hbar\omega_4 \approx U\pm\epsilon,\hspace{5mm}
	\label{eq:approximation}
	U \pm \epsilon \pm \hbar\omega_z \approx U \pm \epsilon, \hspace{5mm} U \pm \epsilon \pm \hbar\delta\omega_z \approx U\pm \epsilon. 
\end{equation}
These are reasonable assumptions, if we exploit the energy hierarchy 
$t^2/(U-\epsilon), \Omega^2/(U-\epsilon)\ll \hbar\omega_z, \hbar\delta \omega_z \ll U - \epsilon$. 
As a consequence, our upcoming results are not valid at very large detunings, when $\epsilon$ is too close to $U$.

\subsection{Monochromatic and bichromatic Rabi frequencies}

We calculated the monochromatic Rabi frequencies using Floquet theory and third-order Schrieffer-Wolff perturbation theory.
We illustrate the calculation of the Rabi frequency of the Q1$^\textrm{P4}$ transition (i.e., the $\ket{\downarrow,\downarrow} \to \ket{\uparrow,\downarrow}$ transition) in Suppl. Fig.~\ref{fig:floquet-levels}b. 
There, the blue horizontal lines represent Floquet levels, i.e., diagonal elements of the Floquet matrix $\mathcal{H}_F$ shown in Eq.~\eqref{Floquet-matrix}.
The green lines depict the off-diagonal matrix elements of the Floquet matrix, which play the role of the perturbation.
(Note that not all Floquet levels and not all perturbation matrix elements are shown in Suppl. Fig.~\ref{fig:floquet-levels}.)
Using Schrieffer-Wolff perturbation theory, we derive an effective $2\times 2$ Floquet matrix $\mathcal{H}_{F,\mathrm{eff}}$ for the relevant subspace of the two Floquet levels highlighted with pink background in Suppl. Fig.~\ref{fig:floquet-levels}b.
The absolute value of the off-diagonal matrix element of the effective Floquet matrix provides the Rabi frequency via
$f_R = 2 \left| \left(\mathcal{H}_{F,\mathrm{eff}}\right)_{1,2} \right|/h$.
This procedure is repeated for all the four single-qubit transitions, and the corresponding Rabi-frequency formulas are shown as the first four entries of the second column of Supplementary Table \ref{tab:frequencies}.

The calculation of the bichromatic Rabi frequencies is similar, the only difference being that it requires fourth order Schrieffer-Wolff transformation, because the leading-order result incorporates two matrix elements of the ac electric fields. 
A corresponding illustrative Floquet level structure and perturbation matrix elements are shown in Fig. \ref{fig:floquet-levels}c. 
Results are shown as the three last entries of the second column of the  Supplementary Table \ref{tab:frequencies}.

\renewcommand{\arraystretch}{1.5}
\begin{table}[H]
	\centering
	\begin{tabular}{|c|c|c|c|} \hline
		Transition & Formula & Measured frequency (MHz) & Calculated frequency (MHz) \\ \hline  
		$\mathrm{Q2^{P2}}$ & $f_R=\frac{4 \epsilon_{\text{P2}}\epsilon U\Omega t}{h(U^2-\epsilon^2)^2}$ & $20.66 \pm 0.05$ & $19.0 \pm 5.1$\\ \hline
		$\mathrm{Q2\_^{P2}}$ & $f_R=\frac{4 \epsilon_{\text{P2}}\epsilon U\Omega t}{h(U^2-\epsilon^2)^2}$ & $22.1 \pm 0.1$ & $18.6 \pm 5.0$\\ \hline
		$\mathrm{Q1^{P4}}$ & $f_R=\frac{4 \epsilon_{\text{P4}}\epsilon U\Omega t}{h(U^2-\epsilon^2)^2}$ & $11.76 \pm 0.04$ & $2.9 \pm 0.8$\\ \hline
		$\mathrm{Q1\_^{P4}}$ & $f_R=\frac{4 \epsilon_{\text{P4}}\epsilon U\Omega t}{h(U^2-\epsilon^2)^2}$ & $5.65 \pm 0.02$ & $3.0 \pm 0.8$\\ \hline
		$\mathrm{Q1^{-P2,P4}}$ & $f_R=\frac{2\epsilon_{\text{P2}}\epsilon_{\text{P4}}U(U^2+3\epsilon^2)\Omega t}{h(U^2-\epsilon^2)^3}$ & $6.24 \pm 0.04$ & $0.59 \pm 0.16$\\ \hline
		$\mathrm{Q2^{-P2,P4}}$ & $f_R=\frac{2\epsilon_{\text{P2}}\epsilon_{\text{P4}}U(U^2+3\epsilon^2)\Omega t}{h(U^2-\epsilon^2)^3}$ & $1.695 \pm 0.008$ & $0.40 \pm 0.11$\\ \hline
		$\mathrm{Q1^{P2,P4}}$ & $f_R=\frac{2\epsilon_{\text{P2}}\epsilon_{\text{P4}}U(U^2+3\epsilon^2)\Omega t}{h(U^2-\epsilon^2)^3}$ & $4.23 \pm 0.01$ & $0.65 \pm 0.17$\\ \hline
	\end{tabular}
	\caption{\textbf{Comparison of measured and calculated monochromatic and bichromatic Rabi frequencies. } Theoretical Rabi frequencies of different monochromatic and bichromatic transitions were calculated with Floquet theory and Schrieffer-Wolff transformation, using the spin-conserving tunneling $t$ and spin-flip tunneling $\Omega$ obtained from fitting the resonance anticrossings. The errors of the calculated frequencies originate from the uncertainties of the average spin-conserving and spin-flip tunneling, see Supplementary Table \ref{Tab:avg_hopping}.}  
	\label{tab:frequencies}
\end{table}
\renewcommand{\arraystretch}{1}

Note that the Rabi-frequency formulas in the second column of Supplementary Table \ref{tab:frequencies} show a high degree of symmetry, i.e. entries 1 and 2 are identical, entries 3 and 4 are identical, and entries 5, 6, 7 are identical. 
This is due to the approximations highlighted around Eq.~\eqref{eq:approximation}.
Without those approximations,  differences between the Rabi-frequency formulas would be retained. Another consequence of the approximations mentioned before is that the bichromatic Rabi frequencies do not depend on the driving frequencies $\omega_2$ and $\omega_4$.

Moreover, Supplementary Table \ref{tab:frequencies} contains the measured and calculated Rabi frequencies for different monochromatic and bichromatic resonances. The  Rabi-frequencies are calculated by the substitution of the following numerical values into the formulas.  The value of the charging energy $U$=2.56 meV is obtained from the experimentally determined value of the lever arm $\alpha=0.0917$ eV/V from photon-assisted tunnelling experiments and from the fitting value Eq. (\ref{eq:fitresult:Ot}). To determine the numerical value of $\epsilon_{\text{P2}}$ and $\epsilon_{\text{P4}}$ we substitute the actual plunger gate voltage at the driving frequency from Fig. \ref{fig:attenuation}b into the definitions after Eq. (\ref{eq:delta_epsilon}).  All of the measurements were carried out at (vP1,vP2)=(-10,10) mV, where $\epsilon$=1.83 meV. 
$\Omega$ and $t$ are obtained from the fitting of the resonance anticrossing, numerical values of  them are listed in Table. \ref{Tab:avg_hopping}.

In the following, we discuss the main observations for the measured and calculated Rabi frequencies.

(1) There is a significant discrepancy between the measured and calculated frequencies in the case of all monochromatic transitions driven by plunger P4 and all bichromatic transitions. We attribute this to the fact that our model contains only an interdot driving effect, the detuning modulation, which is sufficient to understand the anticrossings as well as the detuning dependence of the anticrossing (see section F).
However, our model does not contain an intradot effect, such as g-tensor modulation resonance (g-TMR) 
\cite{kato2003gigahertz,crippa2018electrical}. The interference of g-TMR and detuning modulation can lead to a significant difference between the Rabi-frequencies of $\mathrm{Q1^{P4}}$ and $\mathrm{Q1\_^{P4}}$, and this is what we see from the experiments. Furthermore, the calculated monochromatic Rabi frequencies driven by P2 agree well with the measured frequencies, these suggest that the g-TMR contribution of driving by P2 is negligible to the detuning modulation, but this is not true for the driving by P4. Despite this, we did not take into account the g-TMR mechanism in our theoretical model for two reasons: (i) since the detuning modulation mechanism provides a satisfactory qualitative description of monochromatic and bichromatic resonances, and a quantitative description of resonance anticrossings, (ii) to keep low the number of parameters in the calculations and the fitting process. An additional argument for considering the detuning modulation as the dominant driving mechanism is that direct spin transitions would lead to a Rabi frequency with a significant driving frequency dependence, while the detuning modulation is independent.

(2) Even though in Supplementary Table \ref{tab:frequencies}, the Rabi-frequency formulas for the monochromatic transitions $\mathrm{Q2^{P2}}$ and $\mathrm{Q2\_^{P2}}$ are identical, the calculated Rabi frequencies, which are based on these formulas, are slightly different. 
This is due to a combination of the following two facts.
First, the resonance frequencies of the transitions $\mathrm{Q2^{P2}}$ and $\mathrm{Q2\_^{P2}}$ are different, due to the exchange interaction of the two spins. 
Second, the two Rabi frequencies were measured at the same nominal drive power, but the attenuation of the ac pulses is frequency dependent, hence the drive amplitude reaching the electrons is different at the different Larmor frequencies of $\mathrm{Q2^{P2}}$ and $\mathrm{Q2\_^{P2}}$. 

We note that similarly to monochromatic driving, any single-qubit gate can be realized by appropriately choosing the initial phases ($\varphi_2$ and $\varphi_4$) and durations of the driving pulses in the case of bichromatic EDSR. 
In the co-rotating frame, the qubit rotates around an axis that lies in the x-y plane and forms an angle of $\varphi_2+\varphi_4$ ($\varphi_2-\varphi_4$) with the x-axis if the sum (difference) of the driving frequencies is resonant with the qubit.

\subsection{Qualitative description of resonance anticrossings}

Prominent observations of our experiments are the resonance anticrossings shown in Figs. 3 and 4 of the main text, where we show 5 of them.
In this subsection, we provide a qualitative theoretical description, and a classification of the resonance anticrossings. In the next subsection,  we infer model parameter values by fitting theoretical results on the experimental data.

Each of the 5 resonance anticrossings discussed in the main text is formed at an intersection point of two  resonance lines, where the resonance lines are single or multi-photon transitions between the  ground state $\ket{\downarrow\downarrow}$ and one of the excited states $\ket{\uparrow\downarrow}$, $\ket{\downarrow\uparrow}$, $\ket{\uparrow\uparrow}$. 
Let us consider such an intersection point of two resonance lines, and label the two lines with 1 and 2.
We formulate the resonance conditions corresponding to the two resonance lines as follows:
\begin{equation} \label{eq:reaonance_condition}
	n_{\text{P2}}^{(1)}f_{\text{P2}}+n_{\text{P4}}^{(1)}f_{\text{P4}}=f_{Q_i}, \hspace{5mm} n_{\text{P2}}^{(2)}f_{\text{P2}}+n_{\text{P4}}^{(2)}f_{\text{P4}}=f_{Q_j}.
\end{equation}
Here, 
$f_{Q_i}$ and  $f_{Q_j}$ can be $f_{Q1}$, $f_{Q2}$ or $f_{Q1+Q2\_}$, but $i\neq j$. 
The parameters $n_{\text{P2}}^{(1)}$,
$n_{\text{P4}}^{(1)}$, $n_{\text{P2}}^{(2)}$, and
$n_{\text{P4}}^{(2)}$ are integers, where the upper/lower index labels the resonance line/plunger gate. 
The photon numbers of the two resonance lines (or transitions) are given by $\lvert n_{\text{P2}}^{(1)}\rvert+\lvert n_{\text{P4}}^{(1)}\rvert$ and $\lvert n_{\text{P2}}^{(2)}\rvert+\lvert n_{\text{P4}}^{(2)}\rvert$.
The resonance conditions $\mathrm{Q2^{P2,P4}}$, $\mathrm{Q1^{-P2,P4}}$ and $\mathrm{Q2^{-P2,P4}}$ that can be seen in Fig. 2 in the main text are specific examples of the general expressions in Eq.~\eqref{eq:reaonance_condition}.

\begin{figure}[H]
	\centering
	\includegraphics{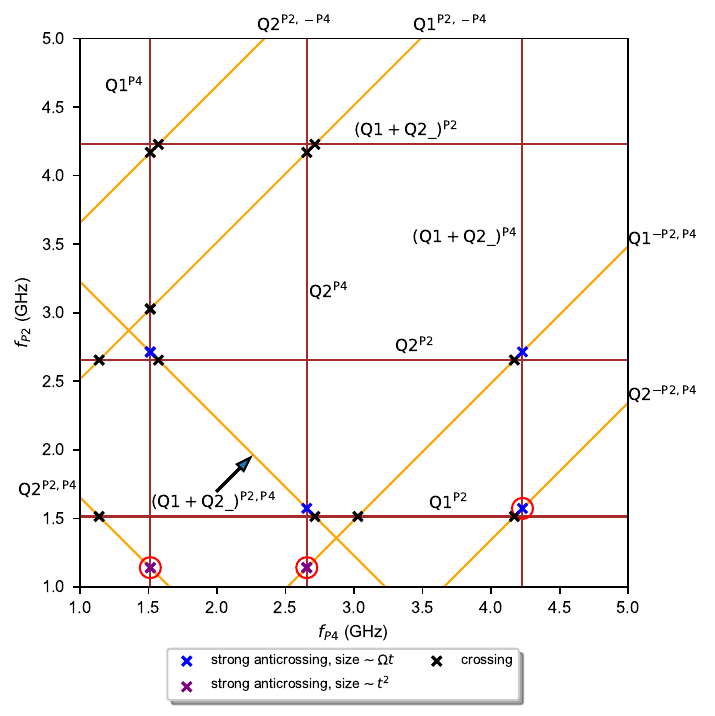}
	
	\caption {\label{fig:map_mono-bi} \textbf{Crossings and anticrossings of single-photon monochromatic and two-photon bichromatic processes.} All possible single-photon monochromatic (brown lines) and two-photon bichromatic processes (yellow lines) are shown which are observable in the shown frequency range. Strong anticrossings mediated by $\Omega t$ tunnelings are shown with blue, while the strong anticrossings mediated by $t^2$ can be seen with purple markers. The intersections which result in crossings are marked with black. The red circles indicate anticrossings which are investigated in detail experimentally and theoretically.}  
\end{figure}

\begin{figure}[H]
	\centering
	\includegraphics{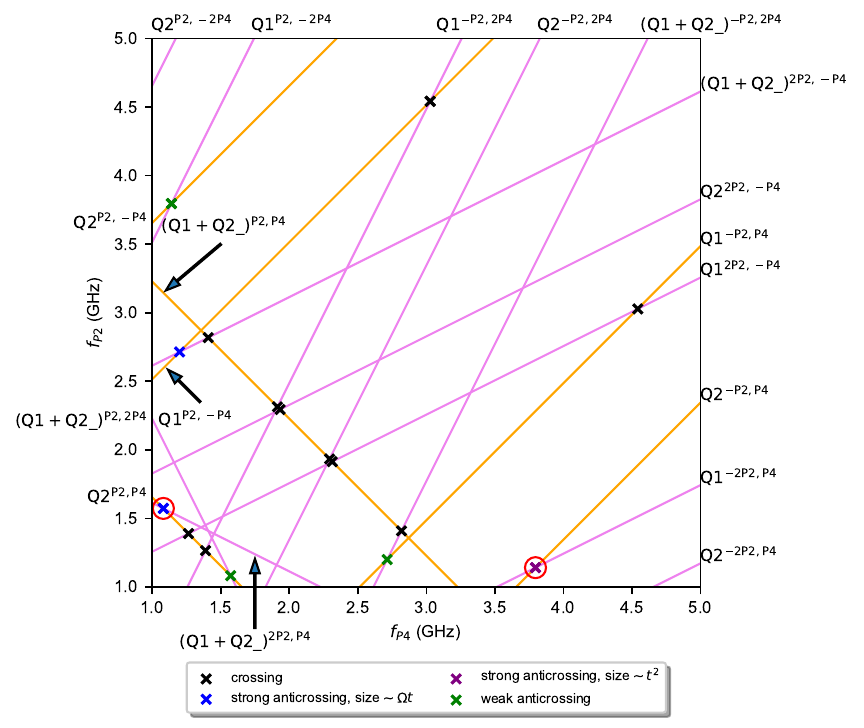}
	
	\caption {\label{fig:map_bi-tri} \textbf{Crossings and anticrossings of two-photon bichromatic and three-photon bichromatic transitions.} All possible two-photon bichromatic (yellow lines) and three-photon bichromatic processes (violet lines) are shown which are observable in the shown frequency range. Strong anticrossings mediated by $\Omega t$ tunnelings are shown with blue, while the strong anticrossings mediated by $t^2$ can be seen with purple markers. The weak anticrossings are denoted by green markers, and the intersections which result in crossings are marked with black. The red circles indicate anticrossings which are investigated in detail experimentally and theoretically. }  
\end{figure}

In Suppl. Fig.~\ref{fig:map_mono-bi}, we plot all single-photon resonance lines as the horizontal and vertical red lines, and all two-photon bichromatic resonance lines as the diagonal orange lines, in the drive frequency range  from 1 GHz to 5 GHz for both $f_{\text{P2}}$ and $f_{\text{P4}}$.
Each intersection point of a single-photon and a two-photon resonance line is marked by a cross ($\times$). 
Three of these intersection points are encircled (red circles).
As we argue below, these three intersection points exhibit strong anticrossings in the experiment, as revealed by the data in the main text.
In a similar fashion, in Suppl. Fig.~\ref{fig:map_bi-tri}, we plot all two-photon bichromatic resonance lines (orange), and all three-photon bichromatic resonance lines (magenta).
Each intersection point of a two-photon and a three-photon resonance line is marked by a cross, and two of the intersection points that appear as resonance anticrossings in the main text (Fig. 3) are encircled (red circles).

At an intersection point of two resonance lines, both resonance conditions of Eq.~\eqref{eq:reaonance_condition} are satisfied, which implies that 
\begin{equation}
	(n_{\text{P2}}^{(1)}-n_{\text{P2}}^{(2)})f_{\text{P2}}+(n_{\text{P4}}^{(1)}-n_{\text{P4}}^{(2)})f_{\text{P4}}=f_{Q_i}-f_{Q_j}.
\end{equation}
In turn, this relation implies that the two excited states of the two resonances are resonantly coupled to each other via a transition with a photon number $\lvert n_{\text{P2}}^{(1)}-n_{\text{P2}}^{(2)} \rvert+\lvert n_{\text{P4}}^{(1)}-n_{\text{P4}}^{(2)} \rvert$.
At such an intersection point, a resonance anticrossing might be formed. 
In fact, we will refer to such an intersection point as a resonance anticrossing, if the strength of the excited-state transition (to be denoted as $|\chi_3|$ below) dominates the strengths of both original resonances 1 and 2 (to be denoted as $|\chi_1|$ and $|\chi_2|$ below).
Such a resonance anticrossing is formed, if each original resonance $k \in \{1,2\}$ satisfies one of the following conditions:
\begin{itemize}
	\item \emph{Condition A.} The photon number of the excited-state transition is smaller than the photon number of the original transition $k$, i.e.,  $\lvert n_{\text{P2}}^{(1)}-n_{\text{P2}}^{(2)} \rvert+\lvert n_{\text{P4}}^{(1)}-n_{\text{P4}}^{(2)} \rvert < \lvert n_{\text{P2}}^{(k)}\rvert+\lvert n_{\text{P4}}^{(k)}\rvert$. 
	\item  \emph{Condition B.} The photon number of the excited-state transition is the same as the photon number of the original transition $k$, i.e., $\lvert n_{\text{P2}}^{(1)}-n_{\text{P2}}^{(2)} \rvert+\lvert n_{\text{P4}}^{(1)}-n_{\text{P4}}^{(2)} \rvert = \lvert n_{\text{P2}}^{(k)}\rvert +\lvert n_{\text{P4}}^{(k)}\rvert$, but the $f_{\text{P2}}$ photon number of the excited-state transition is higher than the $f_{\text{P2}}$ photon number of the original transition $k$, i.e., $\lvert n_{\text{P2}}^{(1)}-n_{\text{P2}}^{(2)} \rvert > \lvert n_{\text{P2}}^{(k)} \rvert$. 
	This condition is based on the fact that the drive amplitude of plunger P2 is much higher than the drive amplitude of plunger P4 in our experiment. 
\end{itemize} 
We now exemplify the general scheme described above.
To this end, we consider a specific resonance anticrossing, labelled as AC1, which appears at the intersection of the monochromatic single-photon resonance $\mathrm{Q1^{P4}}$ and bichromatic two-photon resonance $\mathrm{Q2^{P2,P4}}$. 
For the monochromatic transition, illustrated as the green process in Fig. \ref{fig:floquet-levels}d, the photon numbers are $n_{\text{P2}}^{(1)}=0$ and $n_{\text{P4}}^{(1)}=1$. 
For the bichromatic transition, illustrated as the grey process in Fig. \ref{fig:floquet-levels}d, the photon numbers are $n_{\text{P2}}^{(2)}=1$, $n_{\text{P4}}^{(2)}=1$. 

At the intersection point of the two resonance lines, the transition between the excited states, shown  with red arrows in Fig. \ref{fig:floquet-levels}d, is a single-photon process, where $\lvert n_{\text{P2}}^{(1)}-n_{\text{P2}}^{(2)} \rvert=1$ and $\lvert n_{\text{P4}}^{(1)}-n_{\text{P4}}^{(2)} \rvert=0$. 
This excited-state transition dominates the original bichromatic transition, because the former has a lower photon number than the latter, fulfilling Condition A above. 
Furthermore, the excited-state transition driven by the plunger gate P2 dominates the monochromatic transition that is driven by the weak excitation of plunger gate P4, fulfilling Condition B above.
As seen from the red process in the  Floquet level diagram Fig. \ref{fig:floquet-levels}d, the strength of the excited-state transition, which determines the size of the resonance anticrossing, is proportional to $\frac{\epsilon_{\text{P2}}t^2}{U^2}$. 
(Note that although Suppl. Fig.~\ref{fig:floquet-levels}d illustrates each transition as a single process for simplicity, in fact each transition is defined by the interference of multiple processes.)

The appearance of all 5 experimentally observed resonance anticrossings, denoted by AC1, AC2, $\dots$, AC5, can be explained in the same general framework. 
We list and characterize these resonance anticrossings in  Table \ref{tab:anticrossings1}.
Three of the anticrossings, AC1, AC2, and AC3 appear at the intersection of single-photon and two-photon resonances, whereas two of them, AC4 and AC5, appear at the intersection of two-photon and three-photon resonances. 
We use the notation $\mathrm{Q_i^{n_{\text{P2}}P2,n_{\text{P4}}P4}}$ for the three-photon resonances, 
where $\lvert n_{\text{P2}}\rvert+\lvert n_{\text{P4}}\rvert=3$. 

\renewcommand{\arraystretch}{1.3}
\begin{table}[H]
	\centering
	\begin{tabular}{|c|c|c|c|c|c|} \hline
		Anticrossing & Resonance (1) & ($n_{\text{P2}}^{(1)},n_{\text{P4}}^{(1)}$) & Resonance (2) & ($n_{\text{P2}}^{(2)},n_{\text{P4}}^{(2)}$) & Anticrossing size  \\ \hline  
		AC1 & $\mathrm{Q1^{P4}}$ & (0,1) & $\mathrm{Q2^{P2,P4}}$ & (1,1) & $\propto\frac{\epsilon_{\text{P2}}t^2}{U^2}$ \\ \hline
		AC2 & $\mathrm{Q2^{P4}}$ & (0,1) & $\mathrm{Q1^{-P2,P4}}$ & (-1,1) & $\propto\frac{\epsilon_{\text{P2}}t \Omega}{U^2}$  \\ \hline
		AC3 & $\mathrm{(Q1+Q2\_)^{P4}}$ & (0,1) & $\mathrm{Q2^{-P2,P4}}$ & (-1,1) & $\propto\frac{\epsilon_{\text{P2}}t^2}{U^2}$  \\ \hline
		AC4 & $\mathrm{Q2^{P2,P4}}$ & (1,1) & $\mathrm{(Q1+Q2\_)^{2P2,P4}}$ & (2,1) & $\propto\frac{\epsilon_{\text{P2}}t \Omega}{U^2}$  \\ \hline
		AC5 & $\mathrm{Q2^{-P2,P4}}$ & (-1,1) & $\mathrm{Q1^{-2P2,P4}}$ & (-2,1) & $\propto\frac{\epsilon_{\text{P2}}t^2}{U^2}$ \\ \hline
	\end{tabular}
	\caption{\textbf{Experimentally observed resonance anticrossings.}}
	\label{tab:anticrossings1}
\end{table}
\renewcommand{\arraystretch}{1}

Finally, we classify all intersections of single-, two- and three-photon resonance lines in Table \ref{tab:anticrossings}, as crossings (C), strong anticrossings (SAC), or weak anticrossings (WAC). 
Strong anticrossings are those anticrossings whose size (excited-state transition strength) is proportional to $\frac{\epsilon_{\text{P2}}t\Omega}{U^2}$ or $\frac{\epsilon_{\text{P2}}t^2}{U^2}$. 
Weak anticrossings are those that are not strong.
The size of each weak anticrossing is defined either by a higher-order transition strength, or a transition strength that is governed by $\epsilon_{\text{P4}}$ and not $\epsilon_{\text{P2}}$. 
In Suppl. Figs.~\ref{fig:map_mono-bi} and \ref{fig:map_bi-tri}, we color all intersection points according to this classification.

\renewcommand{\arraystretch}{1.2}
\begin{table}[H]
	\centering
	\begin{tabular}{|c|c|c|c|c|c|} \hline
		\multirow{2}{*}{Transition} & \multirow{2}{*}{1-photon} & 2-photon & 2-photon & 3-photon & 3-photon \\[-2pt] 
		& & (bichromatic) & (monochromatic) & (monochromatic) & (bichromatic) \\ \hline
		1-photon & C & SAC, C & C & C & C \\ \hline
		2-photon & \multirow{2}{*}{} & \multirow{2}{*}{WAC, C} & \multirow{2}{*}{C} & \multirow{2}{*}{C} & \multirow{2}{*}{SAC, WAC, C} \\[-2pt] 
		(bichromatic) & & & & & \\ \hline
		2-photon & \multirow{2}{*}{} & \multirow{2}{*}{} & \multirow{2}{*}{C} & \multirow{2}{*}{C} & \multirow{2}{*}{SAC, WAC, C} \\[-2pt] 
		(monochromatic) & & & & & \\ \hline
		3-photon & \multirow{2}{*}{} & \multirow{2}{*}{} & \multirow{2}{*}{} & \multirow{2}{*}{C} & \multirow{2}{*}{WAC, C} \\[-2pt] 
		(monochromatic) & & & & & \\ \hline
		3-photon & \multirow{2}{*}{} & \multirow{2}{*}{} & \multirow{2}{*}{} & \multirow{2}{*}{} & \multirow{2}{*}{WAC, C} \\[-2pt] 
		(bichromatic) & & & & & \\ \hline
	\end{tabular}
	\caption{\textbf{Anticrossings and crossings of different transitions.}
		The intersection of monochromatic and bichromatic processes with photon numbers from one to three can form strong anticrossings (SAC), weak anticrossings (WAC) and crossings (C). 
	}\label{tab:anticrossings}
\end{table}
\renewcommand{\arraystretch}{1}

\subsection{Quantitative description of resonance anticrossings}

\label{subsec_ftting}

In this subsection, we provide further  details of the quantitative theoretical analysis of the strong resonance anticrossings AC1-AC5 discussed in the main text. 
In particular, we detail how the tunneling parameters $t$ and $\Omega$ have been determined by fitting the theoretical curves of the resonance anticrossing lines on the experimental data. 

\subsubsection{Anticrossing of $\mathrm{Q2^{P2,P4}}$ with $\mathrm{Q1^{P4}}$ $\mathrm{(AC1)}$}

Each anticrossing AC1-AC5 is described by an effective $3\times 3$ Floquet matrix derived from a combination of  Floquet theory and Schrieffer-Wolff perturbation theory, as described in \ref{appsubsec:analytical}.
For AC1, this effective Floquet matrix based on fourth-order Schrieffer-Wolff transformation reads:
\begin{equation}
	\label{anticrossing_Floquet_AC1}
	\mathcal{H}_{\mathcal{F},\mathrm{eff}}=\begin{pmatrix}
		hf_\mathrm{BS,1} & \chi_1 & \chi_2 \\
		\chi_1^* & hf_\mathrm{Q1}+hf_\mathrm{BS,2}-hf_{\text{P4}} & \chi_3 \\ \chi_2^* & \chi_3^* & hf_\mathrm{Q2}+hf_\mathrm{BS,3}-hf_{\text{P2}}-hf_{\text{P4}}
	\end{pmatrix}.
\end{equation}
Here, the Floquet basis states are $\ket{\downarrow\downarrow,0,0}$, $\ket{\uparrow\downarrow,0,-1}$ and $\ket{\downarrow\uparrow,-1,-1}$, $f_\mathrm{BS,1}$ and $f_\mathrm{BS,2}$ are Bloch-Siegert frequencies that can be calculated from Schrieffer-Wolff theory.
Further matrix elements $\chi_1$, $\chi_2$, $\chi_3$ of $\mathcal{H}_{\mathcal{F},\mathrm{eff}}$ are:
\begin{equation}
	\label{anticrossing_Q2sum_curve}   \chi_1=-\dfrac{2\epsilon_{\text{P4}}U\epsilon \Omega \mathrm{e}^{-i\phi_c}t\mathrm{e}^{i\phi}}{\left(U^2-\epsilon^2\right)^2}, \hspace{5mm}
	\chi_2=\dfrac{\epsilon_{\text{P4}}\epsilon_{\text{P2}}\left(U^3+3U\epsilon^2\right)\Omega \mathrm{e}^{-i\phi_c}t\mathrm{e}^{-i\phi}}{\left(U^2-\epsilon^2\right)^3}, \hspace{5mm} \chi_3=\dfrac{2\epsilon_{\text{P2}}U\epsilon\left(t\mathrm{e}^{-i\phi}\right)^2}{\left(U^2-\epsilon^2\right)^2}.
\end{equation}

The $f_\mathrm{BS,1}$, $f_\mathrm{BS,2}$ and $f_\mathrm{BS,3}$ Bloch-Siegert shifts: 
\begin{equation}
	f_\mathrm{BS,1}=-\dfrac{U(U^2+3\epsilon^2)\Omega^2}{h(U^2-\epsilon^2)^3}\epsilon_{\text{P2}}^2(f_{\text{P2}}), \hspace{3mm} f_\mathrm{BS,2}=f_\mathrm{BS,3}=-\dfrac{U(U^2+3\epsilon^2)t^2}{h(U^2-\epsilon^2)^3}\epsilon_{\text{P2}}^2(f_{\text{P2}}),
\end{equation}
where a term proportional to $\epsilon_{\text{P4}}^2$ has been neglected, because the driving by P2 is significantly stronger than the driving by P4. Note that the frequency dependence of the driving has been indicated. The diagonal matrix elements can be shifted by -$f_\mathrm{BS,1}$, this way the $3\times 3$ effective Floquet-matrix becomes: 

\begin{equation}
	\mathcal{H}_{\mathcal{F},\mathrm{eff}}=\begin{pmatrix}
		0 & \chi_1 & \chi_2 \\
		\chi_1^* & hf_\mathrm{Q1}+hf_\mathrm{BS}-hf_{\text{P4}} & \chi_3 \\ \chi_2^* & \chi_3^* & hf_\mathrm{Q2}+hf_\mathrm{BS}-hf_{\text{P2}}-hf_{\text{P4}}
	\end{pmatrix},
\end{equation}
where we introduced the notation:
\begin{equation}
	f_\mathrm{BS}=f_\mathrm{BS,2}-f_\mathrm{BS,1}=-\dfrac{U(U^2+3\epsilon^2)(t^2-\Omega^2)}{h(U^2-\epsilon^2)^3}\epsilon_{\text{P2}}^2(f_{\text{P2}}).
\end{equation}
The appearing term in the numerator $t^2-\Omega^2$ is known from the fit of the spectrum (see Eq. \ref{eq:fitresult:Ot}), so the Bloch-Siegert shift can be calculated.

The power broadening of the resonance $\ket{\downarrow,\downarrow} \leftrightarrow \ket{\uparrow,\downarrow}$ is set by $|\chi_1|$, whereas the power broadening of the resonance $\ket{\downarrow,\downarrow} \leftrightarrow \ket{\downarrow,\uparrow}$ is set by $|\chi_2|$. 
Therefore, a resonance anticrossing can be observed if the third matrix element $\chi_3$, responsible for the strength of the resonance anticrossing, dominates the two power broadenings $\chi_1$ and $\chi_2$. 
We use the term `resonance anticrossing' or `anticrossing' for such cases.
The matrix elements $\chi_1$, $\chi_2$ and $\chi_3$ enlisted in Eq.~\eqref{anticrossing_Q2sum_curve} do fulfill this condition $|\chi_1|, |\chi_2| \ll |\chi_3|$.
First, $\chi_3$ dominates $\chi_2$ as the former (latter) is a 3rd-order (4th-order) perturbative result.
Second, $\chi_3$ dominates $\chi_1$ because the former (latter) is proportional to $\epsilon_{\text{P2}}$ ($\epsilon_{\text{P4}}$), and $\epsilon_{\text{P4}} \ll \epsilon_{\text{P2}}$ holds due to the geometry of the device, see Eq.~\eqref{eq:virtual_plunger}. 

Next, we use the effective Floquet matrix in Eq.~\eqref{anticrossing_Floquet_AC1} to determine the location of the resonance lines in the drive frequency plane.
We focus on the vicinity of the anticrossing, and we consider the perturbative limit $|\chi_1|,|\chi_2| \ll |\chi_3|$. 
In this limit, a decrease in the population of the ground state $\ket{\downarrow,\downarrow}$ is induced by the drive only if the Floquet level $\ket{\downarrow \downarrow,0,0}$ is in resonance with one of the two hybridized states of $\ket{\uparrow \downarrow,0,-1}$ and $\ket{\downarrow \uparrow,-1,-1}$, i.e., when one of the eigenvalues of the lower right $2\times 2$ block of $\mathcal{H}_{\mathcal{F},\mathrm{eff}}$ is zero, i.e., when the determinant of the lower right $2\times 2$ block of $\mathcal{H}_{\mathcal{F},\mathrm{eff}}$ is zero. 
We rewrite this condition in terms of the newly introduced quantities $y_0 = f_{\mathrm{Q1}}$, $x_0=f_\mathrm{Q2}-f_{\text{P2}}-f_{\text{P4}}+\Delta f_{\text{P2}}$, $\Delta f_{\text{P2}}=f_\mathrm{Q2}-f_{\text{P2}}-f_{\text{P4}}$, yielding:
\begin{equation}
	\det 
	\left(
	\begin{array}{cc}
		hy_0+hf_\mathrm{BS} - hf_{\text{P4}} & \chi_3 \\
		\chi_3^* & -h\Delta f_{\text{P2}} +hf_\mathrm{BS}+ hx_0
	\end{array}
	\right) = 0.    
\end{equation}
We derive the shape of the anticrossing resonance lines on the $\Delta f_{\text{P2}}$--$f_{\text{P4}}$ plane from the above determinant condition, by solving the latter for $\Delta f_{\text{P2}}$. 
This yields:
\begin{equation}\label{eq:anticrossing_upperQ2sum}
	\Delta f_{\text{P2}}=x_0-C\epsilon_{\text{P2}}^2(f_\mathrm{Q2}-f_{\text{P4}})-\dfrac{D^2\epsilon_{\text{P2}}^2(f_\mathrm{Q2}-f_{\text{P4}})}{y_0-C\epsilon_{\text{P2}}^2(f_\mathrm{Q2}-f_{\text{P4}})-f_{\text{P4}}}, 
\end{equation}
where the resonance condition $f_{\text{P2}}+f_{\text{P4}}=f_\mathrm{Q2}$ was used to obtain $f_{\text{P2}}$ as a function of $f_{\text{P4}}$, furthermore $C$ and $D$ were introduced: 
\begin{equation}
	C=\dfrac{U(U^2+3\epsilon^2)(t^2-\Omega^2)}{h(U^2-\epsilon^2)^3}, \hspace{3mm} D=\dfrac{2U\epsilon t^2}{h(U^2-\epsilon^2)^2}. 
\end{equation}

Even though this is a single formula, it describes both lines as $f_{\text{P4}}$ can be smaller or greater than $y_0$. 

We fitted Eq.~\eqref{eq:anticrossing_upperQ2sum} on the experimental data; the result is shown in Fig.~ \ref{fig:fitted_anticrossings} \textbf{a.}, the fitting parameters are: 
\begin{equation}
	x_0=-0.569 \hspace{1mm} \mathrm{MHz}, \hspace{5mm}  y_0=1.503 \hspace{1mm} \mathrm{GHz}, \hspace{5mm}    t=20.42  \hspace{1mm} \mu\mathrm{eV}. 
\end{equation}
The frequency dependence of the driving $\epsilon_{\text{P2}}$ was taken into account in the fit by using the measured, filtered and interpolated driving amplitude (see Fig. \ref{fig:attenuation}). 

As an alternative to fitting the anticrossings, the parameter $t$ can also be determined (with higher uncertainty) by measuring the distance between the two branches of the anticrossing and assuming a constant driving amplitude. This distance $d$ is proportional to $\lvert \chi_3 \rvert=hD\epsilon_{\text{P2}}$.
In particular, for AC1, AC2, and AC3, their relation is expressed as $d=2\sqrt{2}\lvert \chi_3 \rvert/h$; for AC4 and AC5, this relation reads: $d\approx 2\sqrt{1+\frac{\sqrt{5}}{5}}\lvert \chi_3 \rvert/h$. 
Note that AC2 is affected by a significant frequency-dependent power attenuation, hence the determination of $\lvert\chi_3\rvert$ by from the distance $d$ is expected to give a result with reduced accuracy.

\subsubsection{Anticrossing of $\mathrm{(Q1+Q2\_)^{2P2,P4}}$ with $\mathrm{Q2^{P2,P4}}$ $\mathrm{(AC4)}$}

The effective Floquet matrix describing this resonance anticrossing AC4, obtained from fourth-order Schrieffer-Wolff transformation, reads: 
\begin{equation}
	\mathcal{H}_{\mathcal{F},\mathrm{eff}}=\begin{pmatrix}
		hf_\mathrm{BS,1} & \chi_1 & \chi_2 \\
		\chi_1^* & hf_\mathrm{Q2}+hf_\mathrm{BS,2}-hf_{\text{P2}}-hf_{\text{P4}} & \chi_3 \\ \chi_2^* & \chi_3^* & hf_\mathrm{Q2}+hf_\mathrm{Q1\_}+hf_\mathrm{BS,3}-2hf_{\text{P2}}-hf_{\text{P4}}
	\end{pmatrix}, 
\end{equation}
where the Floquet basis states in order are $\ket{\downarrow\downarrow,0,0}$, $\ket{\downarrow\uparrow,-1,-1}$ and $\ket{\uparrow\uparrow,-2,-1}$. 
Here, $\chi_2 = 0$, although a fifth-order perturbative calculation reveals that a three-photon process provides a finite $\chi_2$; we neglect that here. 
The other two matrix elements: 
\begin{equation}
	\chi_1=\dfrac{\epsilon_{\text{P4}}\epsilon_{\text{P2}}\left(U^3+3U\epsilon^2\right)\Omega \mathrm{e}^{-i\phi_c}t\mathrm{e}^{-i\phi}}{\left(U^2-\epsilon^2\right)^3}, \hspace{5mm}  \chi_3=\dfrac{2\epsilon_{\text{P2}}U\epsilon \Omega \mathrm{e}^{-i\phi_c}t\mathrm{e}^{i\phi}}{\left(U^2-\epsilon^2\right)^2}.
\end{equation}
The $f_\mathrm{BS,1}$, $f_\mathrm{BS,2}$ and $f_\mathrm{BS,3}$ Bloch-Siegert shifts are: 
\begin{equation}
	f_\mathrm{BS,1}=f_\mathrm{BS,3}=-\dfrac{U(U^2+3\epsilon^2)\Omega^2}{h(U^2-\epsilon^2)^3}\epsilon_{\text{P2}}^2(f_{\text{P2}}), \hspace{3mm} f_\mathrm{BS,2}=-\dfrac{U(U^2+3\epsilon^2)t^2}{h(U^2-\epsilon^2)^3}\epsilon_{\text{P2}}^2(f_{\text{P2}}).
\end{equation}

To determine the resonance anticrossing lines in the $\Delta f_{\text{P2}}$ -- $f_{\text{P4}}$ plane, we follow a procedure analogous to the one in the previous subsection. 
We introduce the following quantities that will serve as the fit parameters: 
$b_0=f_\mathrm{Q2}-f_{\text{P2}}-f_{\text{P4}}+\Delta f_{\text{P2}}$ and  $b_1=f_\mathrm{Q2}+f_\mathrm{Q1\_}-2f_{\text{P2}}-2f_{\text{P4}}+2\Delta f_{\text{P2}}$, and shift the diagonal terms by -$f_\mathrm{BS,1}$, this way we get the following matrix:

\begin{equation}
	\mathcal{H}_{\mathcal{F},\mathrm{eff}}=\begin{pmatrix}
		0 & \chi_1 & \chi_2 \\
		\chi_1^* & hb_0-h\Delta f_{\text{P2}}+hf_\mathrm{BS} & \chi_3 \\ \chi_2^* & \chi_3^* & hb_1+hf_{\text{P4}}-2h\Delta f_{\text{P2}}
	\end{pmatrix},
\end{equation}
where $f_\mathrm{BS}=f_\mathrm{BS,2}-f_\mathrm{BS,1}$.

We obtain the following formula for the lines of the resonance anticrossing:

\begin{equation}\label{eq:anticrossing_lowerQ2sum}
	\Delta f_{\text{P2}}=\dfrac{2a_1+a_2\pm \sqrt{(2a_1-a_2)^2+8\lvert \chi_3\rvert^2/h^2}}{4},     
\end{equation}
where $a_1=b_0+f_\mathrm{BS}$, $a_2=b_1+f_{\text{P4}}$. The $\lvert \chi_3 \rvert$ and $a_1$ parameters depend on $f_{\text{P2}}$ driving frequency, this dependency can be expressed by $f_{\text{P4}}$, $f_{\text{P2}}=f_\mathrm{Q2}-f_{\text{P4}}$, this way the above equation is an explicit expression for $\Delta f_{\text{P2}}$. 

Fitting the experimental data with this formula, we obtain Fig.~ \ref{fig:fitted_anticrossings} \textbf{b}, and the following fitting parameter values: 
\begin{equation}
	b_0=3.011 \hspace{1mm} \mathrm{MHz}, \hspace{5mm}   b_1=-1.064  \hspace{1mm} \mathrm{GHz}, \hspace{5mm}  t\Omega=247.83\hspace{1mm} (\mu\mathrm{eV})^2.
\end{equation}
This value of $t\Omega$ will be used below to determine the hopping energies $t$ and $\Omega$, see Table \ref{tab:t_O_first4}.

\subsubsection{Anticrossing of $\mathrm{Q1^{-2P2,P4}}$ with
	$\mathrm{Q2^{-P2,P4}}$ ($\mathrm{AC5}$)}
The effective Floquet matrix describing this resonance anticrossing reads: 
\begin{equation}
	\mathcal{H}_{\mathcal{F},\mathrm{eff}}=\begin{pmatrix}
		hf_\mathrm{BS,1} & \chi_1 & \chi_2 \\
		\chi_1^* & hf_\mathrm{Q2}+hf_\mathrm{BS,2}+hf_{\text{P2}}-hf_{\text{P4}} & \chi_3 \\ \chi_2^* & \chi_3^* & hf_\mathrm{Q1}+hf_\mathrm{BS,3}+2hf_{\text{P2}}-hf_{\text{P4}}
	\end{pmatrix}, 
\end{equation}
where the Floquet basis states in order are $\ket{\downarrow\downarrow,0,0}$, $\ket{\downarrow\uparrow,1,-1}$ and $\ket{\uparrow\downarrow,2,-1}$. 
Again, the $\chi_2$ matrix element describes a three-photon process, therefore we neglect it. 
The other two off-diagonal matrix elements of the effective Floquet matrix read: 
\begin{equation}
	\chi_1=\dfrac{\epsilon_{\text{P4}}\epsilon_{\text{P2}}\left(U^3+3U\epsilon^2\right)\Omega \mathrm{e}^{-i\phi_c}t\mathrm{e}^{-i\phi}}{\left(U^2-\epsilon^2\right)^3}, \hspace{5mm}  \chi_3=\dfrac{2\epsilon_{\text{P2}}U\epsilon \left(t \mathrm{e}^{i\phi}\right)^2}{\left(U^2-\epsilon^2\right)^2}.
\end{equation}
The $f_\mathrm{BS,1}$, $f_\mathrm{BS,2}$ and $f_\mathrm{BS,3}$ Bloch-Siegert shifts are: 
\begin{equation}
	f_\mathrm{BS,1}=-\dfrac{U(U^2+3\epsilon^2)\Omega^2}{h(U^2-\epsilon^2)^3}\epsilon_{\text{P2}}^2(f_{\text{P2}}), \hspace{3mm}f_\mathrm{BS,2}=f_\mathrm{BS,3}=-\dfrac{U(U^2+3\epsilon^2)t^2}{h(U^2-\epsilon^2)^3}\epsilon_{\text{P2}}^2(f_{\text{P2}}).
\end{equation}

To determine the resonance anticrossing lines in the $\Delta f_{\text{P2}}$--$f_{\text{P4}}$ plane, we follow the procedure as above.
We introduce the following quantities that will serve as the fit parameters: 
$b_0=f_\mathrm{Q2}+f_{\text{P2}}-f_{\text{P4}}-\Delta f_{\text{P2}}$ and $b_1=f_\mathrm{Q1}+2f_{\text{P2}}-2f_{\text{P4}}-2\Delta f_{\text{P2}}$, furthermore we shift the diagonal matrix elements by -$f_\mathrm{BS,1}$:  
\begin{equation}
	\mathcal{H}_{\mathcal{F},\mathrm{eff}}=\begin{pmatrix}
		0 & \chi_1 & \chi_2 \\
		\chi_1^* & hb_0+hf_\mathrm{BS}+h\Delta f_{\text{P2}} & \chi_3 \\ \chi_2^* & \chi_3^* & hb_1+hf_\mathrm{BS}+hf_{\text{P4}}+2h\Delta f_{\text{P2}}
	\end{pmatrix}, 
\end{equation}
where $f_\mathrm{BS}=f_\mathrm{BS,2}-f_\mathrm{BS,1}$. 
We obtain the following formula for the two lines of the resonance anticrossing: 
\begin{equation}\label{eq:anticrossing_lowerQ2diff}
	\Delta f_{\text{P2}}=\dfrac{-(2a_1+a_2)\pm \sqrt{(2a_1-a_2)^2+8\lvert \chi_3 \rvert^2/h^2 }}{4},
\end{equation}
where $a_1=b_0+f_\mathrm{BS}$ and $a_2=b_1+f_{\text{P4}}+f_\mathrm{BS}$. The $\lvert\chi_3\rvert$, $a_1$ and $a_2$ parameters depend on $f_{\text{P2}}$ frequency, which can be expressed with $f_{\text{P4}}$ in order to get an explicit expression for $\Delta f_{\text{P2}}$, $f_{\text{P2}}=f_{\text{P4}}-f_\mathrm{Q2}$. 

Fitting the experimental data with this formula, we obtain Suppl. Fig.~\ref{fig:fitted_anticrossings} \textbf{c.}, and the following fitting parameter values:
\begin{equation}
	b_0=1.046 \hspace{1mm} \mathrm{MHz}, \hspace{5mm}    b_1=-3.795 \hspace{1mm} \mathrm{GHz}, \hspace{5mm}     t=21.26\hspace{1mm} \mu\mathrm{eV}.
\end{equation}
This value of $t$ will be used below to determine the spin-flip tunneling $\Omega$, see Tab. \ref{tab:t_O_first4}.

\subsubsection{Anticrossing of $\mathrm{Q2^{-P2,P4}}$ with $\mathrm{(Q1+Q2\_)^{P4}}$ $\mathrm{(AC3)}$}

\begin{figure}[H]
	\centering
	\includegraphics[width=\columnwidth]{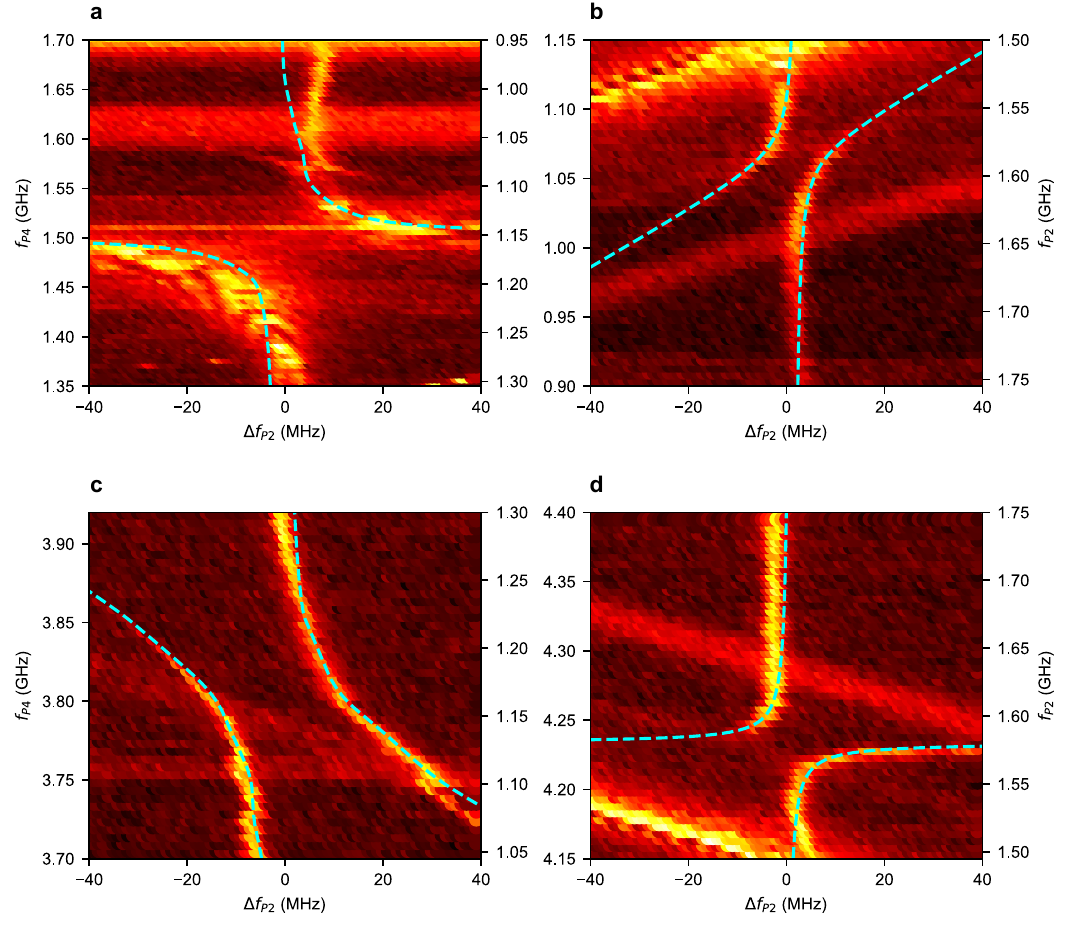}
	\caption {\label{fig:fitted_anticrossings} \textbf{Resonance anticrossings: experiment and theory.} Experimental data is identical to that of Fig. 3 of the main text. \textbf{a.} Resonance anticrossing AC1, experimental data and the fitted theoretical resonance curve (dashed) described by Eq.~\eqref{eq:anticrossing_upperQ2sum}.\textbf{b.} The fit of Eq.~\eqref{eq:anticrossing_lowerQ2sum} on resonance anticrossing AC4. \textbf{c.} The fit of Eq.~\eqref{eq:anticrossing_lowerQ2diff} on resonance anticrossing AC5. \textbf{d.} The fit of Eq.~\eqref{eq:anticrossing_upperQ2diff} on resonance anticrossing AC3.}  
\end{figure}

The effective Floquet matrix describing this resonance anticrossing reads: 
\begin{equation}    \mathcal{H}_{\mathcal{F},\mathrm{eff}}=\begin{pmatrix}
		hf_\mathrm{BS,1} & \chi_1 & \chi_2 \\
		\chi_1^* & hf_\mathrm{Q1}+hf_\mathrm{Q2\_}+hf_\mathrm{BS,2}-hf_{\text{P4}} & \chi_3 \\ \chi_2^* & \chi_3^* & hf_\mathrm{Q2}+hf_\mathrm{BS,3}+hf_{\text{P2}}-hf_{\text{P4}}
	\end{pmatrix}, 
\end{equation}
where the Floquet basis states in order are $\ket{\downarrow\downarrow,0,0}$, $\ket{\uparrow\uparrow,0,-1}$ and $\ket{\downarrow\uparrow,1,-1}$. 
The off-diagonal matrix elements are expressed as: 
\begin{equation}
	\chi_1=-\dfrac{2\epsilon_{\text{P4}}U\epsilon \left(\Omega \mathrm{e}^{-i\phi_c}\right)^2}{\left(U^2-\epsilon^2\right)^2}, \hspace{5mm} 
	\chi_2=\dfrac{\epsilon_{\text{P4}}\epsilon_{\text{P2}}\left(U^3+3U\epsilon^2\right)\Omega \mathrm{e}^{-i\phi_c}t\mathrm{e}^{-i\phi}}{\left(U^2-\epsilon^2\right)^3}, \hspace{5mm} \chi_3=\dfrac{2\epsilon_{\text{P2}}U\epsilon \Omega \mathrm{e}^{i\phi_c} t\mathrm{e}^{-i\phi}}{\left(U^2-\epsilon^2\right)^2}.
\end{equation}
The $f_\mathrm{BS,1}$, $f_\mathrm{BS,2}$ and $f_\mathrm{BS,3}$ Bloch-Siegert shifts are: 
\begin{equation}
	f_\mathrm{BS,1}=f_\mathrm{BS,2}=-\dfrac{U(U^2+3\epsilon^2)\Omega^2}{h(U^2-\epsilon^2)^3}\epsilon_{\text{P2}}^2(f_{\text{P2}}), \hspace{3mm} f_\mathrm{BS,3}=-\dfrac{U(U^2+3\epsilon^2)t^2}{h(U^2-\epsilon^2)^3}\epsilon_{\text{P2}}^2(f_{\text{P2}}).
\end{equation}
To determine the resonance anticrossing lines in the $\Delta f_{\text{P2}}$ -- $f_{\text{P4}}$ plane, we follow the procedure as above.
We introduce the following quantities that will serve as the fit parameters: $x_0=\Delta f_{\text{P2}}-f_\mathrm{Q2}-f_{\text{P2}}+f_{\text{P4}}$ and $y_0=f_\mathrm{Q1}+f_\mathrm{Q2\_}$.
By shifting the diagonal matrix elements by -$f_\mathrm{BS,1}$ we obtain the following matrix: 

\begin{equation}    \mathcal{H}_{\mathcal{F},\mathrm{eff}}=\begin{pmatrix}
		0 & \chi_1 & \chi_2 \\
		\chi_1^* & hy_0-hf_{\text{P4}} & \chi_3 \\ \chi_2^* & \chi_3^* & h\Delta f_{\text{P2}}-hx_0+hf_\mathrm{BS}
	\end{pmatrix}, 
\end{equation}
where $f_\mathrm{BS}=f_\mathrm{BS,2}-f_\mathrm{BS,1}$. From this matrix the equation of the resonance curve: 
\begin{equation}\label{eq:anticrossing_upperQ2diff}
	\Delta f_{\text{P2}}=\dfrac{D^2\epsilon_{\text{P2}}^2(f_{\text{P4}}-f_\mathrm{Q2})}{y_0-f_{\text{P4}}}+x_0+C\epsilon_{\text{P2}}^2(f_{\text{P4}}-f_\mathrm{Q2}),
\end{equation}
where $C$ and $D$ parameters are: 
\begin{equation}
	C=\dfrac{U(U^2+3\epsilon^2)(t^2-\Omega^2)}{h(U^2-\epsilon^2)^3}, \hspace{3mm} D=\dfrac{2U\epsilon t\Omega}{h(U^2-\epsilon^2)^2}. 
\end{equation}
Fitting the experimental data with this formula, we obtain Suppl. Fig.~\ref{fig:fitted_anticrossings} \textbf{d.}, and the following fitting parameter values: 
\begin{equation}
	x_0=-0.6716 \hspace{1mm} \mathrm{MHz}, \hspace{5mm} y_0=4.2335  \hspace{1mm} \mathrm{GHz}, \hspace{5mm} t\Omega=250.78 \hspace{1mm} (\mu\mathrm{eV})^2.
\end{equation}

\subsubsection{Anticrossing of $\mathrm{Q1^{-P2,P4}}$ with $\mathrm{Q2^{P4}}$ $\mathrm{(AC2)}$}

The resonance anticrossing AC2 is measured at five different detuning values from (-11,11) mV to (-7,7) mV virtual plunger voltages, as shown in Fig. 4. 
Spin dynamics in the vicinity of this resonance anticrossing is affected by a strongly frequency-dependent power attenuation.
In fact, the drive signal delivered by plunger gate P2 has a strong low-frequency cutoff, when $f_{\text{P2}}$ is below approximately 1.05 GHz, see Fig. \ref{fig:attenuation} b. i. 
We take into account this effect by using the measured driving amplitude (see Fig. \ref{fig:attenuation}). 

The effective Floquet matrix describing this resonance anticrossing reads:
\begin{equation}\label{}
	\mathcal{H}_{\mathcal{F},\mathrm{eff}}=\begin{pmatrix}
		hf_\mathrm{BS,1} & \chi_1 & \chi_2 \\ 
		\chi_1^* & hf_\mathrm{Q2}+hf_\mathrm{BS,2}-hf_{\text{P4}} & \chi_3 \\
		\chi_2^* & \chi_3^* & hf_\mathrm{Q1}+hf_\mathrm{BS,3}+hf_{\text{P2}}-hf_{\text{P4}}
	\end{pmatrix},
\end{equation}
where the Floquet basis states in order are $\ket{\downarrow\downarrow,0,0}$, $\ket{\downarrow\uparrow,0,-1}$ and $\ket{\uparrow\downarrow,1,-1}$. 
The off-diagonal matrix elements are expressed as: 
\begin{equation}\label{eq:Q1diff_chi}
	\chi_1=\dfrac{2\epsilon_{\text{P4}}U\epsilon \Omega \mathrm{e}^{-i\phi_c}t\mathrm{e}^{-i\phi}}{\left(U^2-\epsilon^2\right)^2}, \hspace{5mm}
	\chi_2=-\dfrac{\epsilon_{\text{P4}}\epsilon_{\text{P2}}\left(U^3+3U\epsilon^2\right)\Omega \mathrm{e}^{-i\phi_c}t\mathrm{e}^{i\phi}}{\left(U^2-\epsilon^2\right)^3}, \hspace{5mm} \chi_3=\dfrac{2\epsilon_{\text{P2}}U\epsilon\left(t\mathrm{e}^{i\phi}\right)^2}{\left(U^2-\epsilon^2\right)^2}.
\end{equation}

The Bloch-Siegert frequency shifts $f_\mathrm{BS,1}$, $f_\mathrm{BS,2}$ and $f_\mathrm{BS,3}$ are: 

\begin{equation}
	f_\mathrm{BS,1}=-\dfrac{U(U^2+3\epsilon^2)\Omega^2}{h(U^2-\epsilon^2)^3}\epsilon_{\text{P2}}^2(f_{\text{P2}}), \hspace{3mm} f_\mathrm{BS,2}=f_\mathrm{BS,3}=-\dfrac{U(U^2+3\epsilon^2)t^2}{h(U^2-\epsilon^2)^3}\epsilon_{\text{P2}}^2(f_{\text{P2}}).
\end{equation}

By introducing $x_0=f_{\text{P4}}-f_{\text{P2}}-f_{Q1}$ and $y_0=f_\mathrm{Q2}$ parameters and shifting the diagonal elements by -$f_\mathrm{BS,1}$, the relevant $2\times 2$ subspace of $\mathcal{H}_{\mathcal{F},\mathrm{eff}}$ and the condition for an anticrossing becomes: 

\begin{equation}
	\det 
	\left(
	\begin{array}{cc}
		hy_0 - hf_{\text{P4}} +hf_\mathrm{BS} & \chi_3 \\
		\chi_3^* & h\Delta f_{\text{P2}} - hx_0 + hf_\mathrm{BS}
	\end{array}
	\right) = 0,    
\end{equation}
where $f_\mathrm{BS}=f_\mathrm{BS,2}-f_\mathrm{BS,1}$. 
From this condition $\Delta f_{\text{P2}}$ can be expressed: 

\begin{equation}\label{eq:AC2_line}
	\Delta f_{\text{P2}}=\dfrac{D^2\epsilon_{\text{P2}}^2(f_{\text{P4}}-f_\mathrm{Q1})}{y_0-f_{\text{P4}}-C\epsilon_{\text{P2}}^2(f_{\text{P4}}-f_\mathrm{Q1})}+x_0+C\epsilon_{\text{P2}}^2(f_{\text{P4}}-f_\mathrm{Q1}),
\end{equation}
where the $C$ and $D$ parameters: 
\begin{equation}
	C=\dfrac{U(U^2+3\epsilon^2)(t^2-\Omega^2)}{h(U^2-\epsilon^2)^3}, \hspace{3mm} D=\dfrac{2U\epsilon t^2}{h(U^2-\epsilon^2)^2}.
\end{equation}

\begin{figure}[H]
	\centering
	\includegraphics[width=\columnwidth]{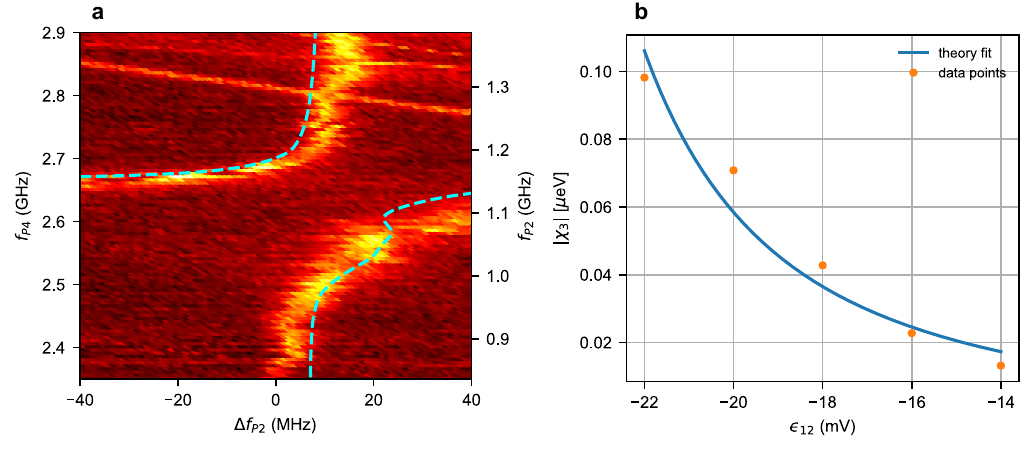}
	
	\caption {\label{fig:anticrossing_detuning} \textbf{Detuning dependence of anticrossing AC2.} \textbf{a.} Anticrossing AC2 at $\epsilon_{12}=-22$ mV point and the fitted theoretical resonance line Eq. \ref{eq:AC2_line} \textbf{b.} The absolute values of $\chi_3$ calculated at a constant driving ($\epsilon_{\text{P2}}(f_{\text{P2}}=1.1$ GHz)) as a function of detuning voltage $\epsilon_{12}$ ranging from -22 mV to -14 mV with orange markers. The absolute value of $\chi_3$ (see Eq. \ref{eq:Q1diff_chi}) is fitted on the datapoints, the result of the fit can be seen with blue line.}  
\end{figure}

We fit the five replicas of resonance anticrossing AC2 measured at different detunings, from $\epsilon_{12}=-14$ mV to $\epsilon_{12}=-22$ mV.  The fitting of the theoretical curves at $\epsilon_{12}=-22$ mV point can be seen in Fig. \ref{fig:anticrossing_detuning} \textbf{a.}, while the resulting parameters from the five fits of the anticrossing at different detunings can be found in Tab. \ref{tab:Q1diff_parameters}. 

\renewcommand{\arraystretch}{1.3}
\begin{table}[H]
	\centering
	\begin{tabular}{|c|c|c|c|} \hline
		$\epsilon_{12}$ (mV)  & $x_0$ (MHz) & $y_0$ (GHz) & $t$ ($\mu$eV) \\ \hline  
		-14 & -7.420 & 2.617 & 14.99 \\ \hline
		-16 & -5.585 & 2.628 & 16.54 \\ \hline
		-18 & -0.899 & 2.644 & 18.59\\ \hline
		-20 & 6.574 & 2.652 & 18.91 \\ \hline
		-22 & 7.035 & 2.666 & 16.53 \\ \hline
	\end{tabular}
	\caption{\textbf{Fitting parameters of anticrossing AC2. 
	}}\label{tab:Q1diff_parameters}
\end{table}
\renewcommand{\arraystretch}{1}

In the case of constant power the $\lvert \chi_3 \rvert$ is proportional to the size of the anticrossing, therefore the observed dependence of the size on the virtual plunger gate voltages is captured by
Eq.~\eqref{eq:Q1diff_chi}. For the AC2 anticrossings the results of the fits yield the $t$ hopping parameters, which are slightly different for different detunings. Using those values the $\chi_3$ parameters can be calculated and the theoretical formula can be fitted on the data points. This can be seen in Fig. \ref{fig:anticrossing_detuning} b, the resulting $t$ value from the fit $t=17.2$ $\mu$eV.  

Using the values obtained from the fits ($t$ or $t\Omega$) and the relation between $t$ and $\Omega$
Eq.~\eqref{eq:fitresult:Ot}, we calculate the spin-conserving and spin-flip tunnelings $t$ and $\Omega$. The calculated parameters can be found in Tab. \ref{tab:t_O_first4}. 

\renewcommand{\arraystretch}{1.3}
\begin{table}[H]
	\centering
	\begin{tabular}{|c|c|c|c|} \hline
		Anticrossing & $\epsilon_{12}$ (mV) & $t$ ($\mu$eV) & $\Omega$ ($\mu$eV) \\ \hline  
		AC1  & -20 & 20.42 & 17.19\\ \hline
		AC2  & -22 & 16.53 & 12.32\\ \hline
		AC2  & -20 & 18.91 & 15.37\\ \hline
		AC2  & -18 & 18.59 & 14.98\\ \hline
		AC2  & -16 & 16.54 & 12.34\\ \hline
		AC2  & -14 & 14.99 & 10.18\\ \hline
		AC3  & -20 & 17.85 & 14.05\\ \hline
		AC4  & -20 & 17.77 & 13.95\\ \hline
		AC5  & -20 & 21.26 & 18.18\\ \hline
	\end{tabular}
	\caption{\textbf{Calculated spin-conserving and spin-flip tunnelings using the different fitted anticrossings.} 
		Using Eq.~\eqref{eq:fitresult:Ot}, the relation between $t$ and $\Omega$ the tunneling amplitudes can be calculated from the fit results of the anticrossings.   
	}\label{tab:t_O_first4}
\end{table}
\renewcommand{\arraystretch}{1}

We take the average of the different $t$ and $\Omega$ values from Tab. \ref{tab:t_O_first4}, these average hopping parameters can be found in Tab. \ref{tab:avg_hopping}, and are used to plot the anticrossings in Figs. 3 and 4, using the offset parameters obtained from the fits ($x_0$ and $y_0$ or $b_0$ and $b_1$.). In Fig. 3, the four anticrossings are plotted using average offset parameters, the average of $x_0$ and $b_0$ ($x_0$ corresponding to AC3 and $b_0$ to AC5) parameters is used to plot AC3 and AC5, similarly for AC1 and AC4.

\renewcommand{\arraystretch}{1.3}
\begin{table}[H]
	\centering
	\begin{tabular}{|c|c|} \hline
		Average $t$ [$\mu$eV] & Average $\Omega$ [$\mu$eV] \\ \hline  
		$18.1\pm 1.9$& $14.3 \pm 2.4$\\ \hline
	\end{tabular}
	\caption{\label{Tab:avg_hopping}\textbf{Average hopping parameters.} An average $t$ and $\Omega$ value was calculated using the values from Tab. \ref{tab:t_O_first4}.}\label{tab:avg_hopping}
\end{table}
\renewcommand{\arraystretch}{1}

%